\RequirePackage{fix-cm}
\documentclass[smallextended]{svjour3}       
\smartqed  
\usepackage{graphicx}
\usepackage{enumerate}
\usepackage{cite}
\usepackage{enumerate}

\usepackage{amsthm,amsmath,amssymb,mathtools,newtxmath,bbm} 
\usepackage{xcolor}

\begin{document}

\title{Game-Theoretic Frameworks for Epidemic Spreading and Human Decision-making: A Review 
}


\author{Yunhan Huang         \and
        Quanyan Zhu 
}


\institute{Y. Huang and Q. Zhu \at
              370 Jay Street, Brooklyn, New York, United States \\
              \email{\{yh.huang, qz494\}@nyu.com}           
}

\date{Received: date / Accepted: date}

\maketitle

\begin{abstract}
This review presents and reviews various solved and open problems in developing, analyzing, and mitigating epidemic spreading processes under human decision-making. We provide a review of a range of epidemic models and explain the pros and cons of different epidemic models. We exhibit the art of coupling epidemic models and decision models in the existing literature. {More specifically, we provide answers to fundamental questions in human decision-making amid epidemics, including what interventions to take to combat the disease, who are decision-makers, when to take interventions, and how to make interventions.}

Among many decision models, game-theoretic models have become increasingly crucial in modeling human responses/behavior amid epidemics in the last decade.  In this review, we motivate the game-theoretic approach to human decision-making amid epidemics.
This review provides an overview of the existing literature by developing a multi-dimensional taxonomy, which categorizes existing literature based on multiple dimensions, including 1) types of games, such as differential games, stochastic games, evolutionary games, and static games; 2) types of interventions, such as social distancing, vaccination, quarantine, taking antidotes, etc.; 3) the types of decision-makers, such as individuals, adversaries, and central authorities at different hierarchical levels. A fine-grained dynamic game framework is proposed to capture the essence of game-theoretic decision-making amid epidemics. We showcase three representative {frameworks} with unique ways of integrating game-theoretic decision-making into the epidemic models from a vast body of literature. {Each of the three framework has a unique way of modeling, conducting analytical analysis, and deriving results.} In the end, we identify several main open problems and research gaps left to be addressed and filled.

\keywords{Dynamic Games \and Stochastic Games \and Infectious Diseases \and Epidemic Spreading \and Human-in-the-Loop Systems \and COVID-19}
\end{abstract}

\setcounter{secnumdepth}{3}
\setcounter{tocdepth}{3}
\tableofcontents

\section{Introduction}\label{intro}

The advancement of Information and Communication Technologies (ICTs) and transportation technologies has been connecting entities, including people and devices, in various ways and improving the quality of lives globally. While benefiting from increasing connectivity, we have also experienced several toxic ``side effects'' that are bought by hidden spreading processes over the underlying networks. Such processes include the spread of contagious diseases over human contact networks and animal populations, the diffusion of viruses or worms over communication and computer networks, and the propagation of rumors/fake news over social media. The recent COVID-19 pandemic, which has been taking a devastating toll on the physical and economic well-being of people across the world, needs more earnest heed to these processes. Certainly, a fundamental understanding of the evolution and control of these processes will contribute to alleviating the threats to the safety, well-being, and security of people and other interconnected systems around the world. 

The underlying networks on which these processes spread are usually large-scale complex networks composed of intelligent and strategic individuals with different beliefs, perceptions, and objectives. The scale and complexity of the underlying networks, the unpredictability of individuals' behavior, and the unavailability of accurate and timely data pose challenges to the fundamental understanding of the evolution and control of these processes.

The inclusion of human decision-makers in the epidemic spreading process creates challenges. It becomes a significant hurdle for researchers to understand human-in-the-loop spreading processes at a fundamental level. The key to clear this hurdle is to integrate decision models into epidemic models. Mathematical models of epidemic spreading started 200 years ago by Daniel Bernoulli \cite{bernoulli1760essai}, at the dawn of the industrial revolution. Many papers have been dedicated to the modeling of epidemic spreading over the last 200 years. But not until very recent years has there been studies investigating human decision-making amid epidemics and the effect of human responses on the spreading processes. In this review, we review and present various solved and open problems in developing, analyzing, and mitigating  the epidemic spreading process with human decision-making. We provide a tutorial on epidemic models and the pros and cons of different epidemic models. We explain in detail how decision models are introduced and integrated into epidemic models in the existing literature. For example, we provide concrete examples regarding what interventions can be taken by individuals and the central authority to fight against the epidemic, when interventions are taken, and how interventions are modeled.

Among various decision models, game-theoretic models have become prominent in modeling human responses/behavior amid epidemics in the last decade. The popularity of game-theoretic models for human-in-the-loop epidemics is primarily due to the following reasons. One rationale  is the large-scale nature of the human population.  Centralized decision-making becomes intractable in a large-scale network. Game-theoretic models provide a bottom-up decentralized modeling framework that naturally makes the computation and design scalable. The second one is that individuals living amid an epidemic may not be willing to comply with the suggested protocols, and individuals are mostly self-interested. The third one is that game theory as a mature and extensive field offers a set of relevant concepts and analytical techniques that can be leveraged to study human behavior amid epidemics. In this review, we demonstrate that game-theoretic frameworks are powerful in modeling the spreading processes of human-in-the-loop epidemics. We provide a multi-dimensional taxonomy of the existing literature that has proposed, studied, and analyzed game-theoretic models for human-in-the-loop epidemics. Among existing literature, we showcase three representative frameworks with unique ways of integrating game-theoretic decision-making into the epidemic models. The uniqueness of each of these three frameworks distinguishes them from each other by means of their models, analytical methods, and results. 

Despite a recent surge in the literature about game-theoretic models for studying the human-in-the-loop epidemic spreading, a number of open problems and research gaps are left to be addressed and filled. Hence, we use one section to discuss emerging topics. 
As we continue to witness the devastating toll of the pandemic on human society, this review aims to introduce to more researchers, especially game and dynamic game theorists the subject of game-theoretic modeling of human-in-the-loop epidemic spreading processes. Their contributions to understanding and mitigating the human-in-the-loop epidemic spreading would make a significant societal impact.

\subsection{Mathematical Preliminaries}

\textit{Graph Theory}: A directed graph (network) is a pair $\mathcal{G} = (\mathcal{N},\mathcal{E})$, where $\mathcal{N}$ is the set of nodes representing individuals involved and $\mathcal{E}\subset \mathcal{N}\times \mathcal{N}$ is the set of edges representing connections between two individuals. The size of the network is $N=\vert \mathcal{N}\vert$.  Given $\mathcal{G}$, an edge from node $i\in\mathcal{N}$
to node $j\in\mathcal{N}$ is denoted by $(i,j)$. When $(i,j)\in\mathcal{E}$ implies $(j,i)\in\mathcal{E}$ and vice versa, the graph $\mathcal{G}$ is undirected. {For an undirected network, we denote $\mathcal{N}_i = \{j|(i,j)\in\mathcal{E}\}$ the set of neighbors of node $i$.} For a directed network, we denote $N_i^{out}$ the out-neighbor set of individual $i$, which is defined as $\mathcal{N}_i = \{j|(j,i)\in\mathcal{E}\}$. Let $A\in\mathbb{R}^{N \times N}$ {be} the adjacency matrix for {an} unweighted graph $\mathcal{G}$ with elements $a_{ij}=1$ if and only if $(1,j)\in\mathcal{E}$. Otherwise, $a_{ij}=0$. For an unweighted network, we use $W$ as the adjacency matrix with elements $w_{ij}\in\mathbb{R}_+$. 

The number of neighbors a node has {is called the degree of a node}. Given a graph $\mathcal{G}$, $P(k)$ is the proportion of nodes who have degree $k$. The average degree of a graph is denoted by $\langle  k \rangle \coloneqq \sum_{k} k P(k)$. {We define $\langle k^2 \rangle \coloneqq \sum_k k^2 P(k)$.} If $\mathcal{G}$ is a scale-free graph,  its degree distribution follows $P(k)\sim k^{-\gamma}$ with $\gamma$ ranges from $2$ to $3$.  If $\mathcal{G}$ is a regular graph, then each node in $\mathcal{G}$ has the same degree.

\textit{Notation:} 
For a square matrix $M$, $\lambda_{max}(M)$ is the spectral radius of matrix $M$. Given a vector $(x_1,x_2,\cdots, x_N)$, $\textrm{diag}(x_1,x_2,\cdots,x_N)$ is an $N\times N$ matrix whose elements on the $i$-th row and the $i$-th column is $x_i$. The identity matrix is denoted by $I_d$. We use $\mathbb{E}[\cdot]$ and $\mathbb{P}(\cdot)$ to denote the expected value and the probability of the argument. Given $\mathbf{x}= (x_1,x_2,\cdots, x_N)$, $\mathbf{x}_{-i}$ denotes the vector {$\mathbf{x}$ with element $x_i$ removed.} Given a graph $\mathcal{G}$ and a vector associated with its nodes $\mathbf{x}= (x_1,x_2,\cdots, x_N)$, $\mathbf{x}_{\mathcal{N}_i}$ denotes the vector that includes all the elements associated with node $i$ and its neighbors. Let $\Delta^n$ be the probability simplex of dimension $n$, i.e., {$\Delta^n = \{x\in\mathbb{R}^n \vert x_1 + x_2 + \cdots+ x_n = 1, x_i \geq 0, \forall i=1,2,\cdots,n\}$}.

\section{Epidemic Models \& Decision Models}\label{EpidemicAndDecision}

To understand the evolution and control of epidemic spreading at a fundamental level, we need to understand both the epidemic model and the decision model, and the art of coupling these two models. 

\begin{figure}
  \includegraphics[width=1\textwidth]{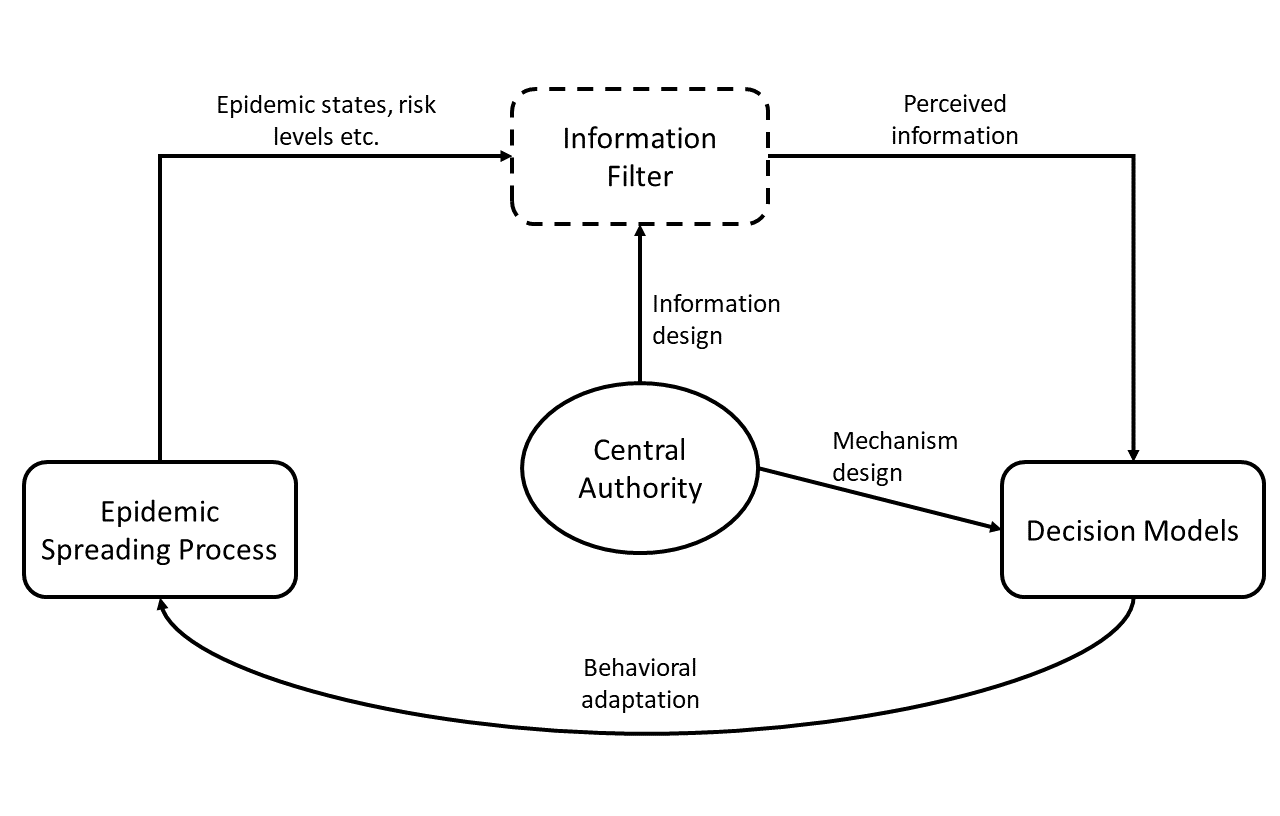}\vspace{-0.3in}
\caption{Schematic illustration of the human-in-the-loop epidemic framework consisting of epidemic processes and decision models.}
\label{fig:EpidemicDecisionLoop}       
\end{figure}
%
\begin{figure*}
  \includegraphics[width=1\textwidth]{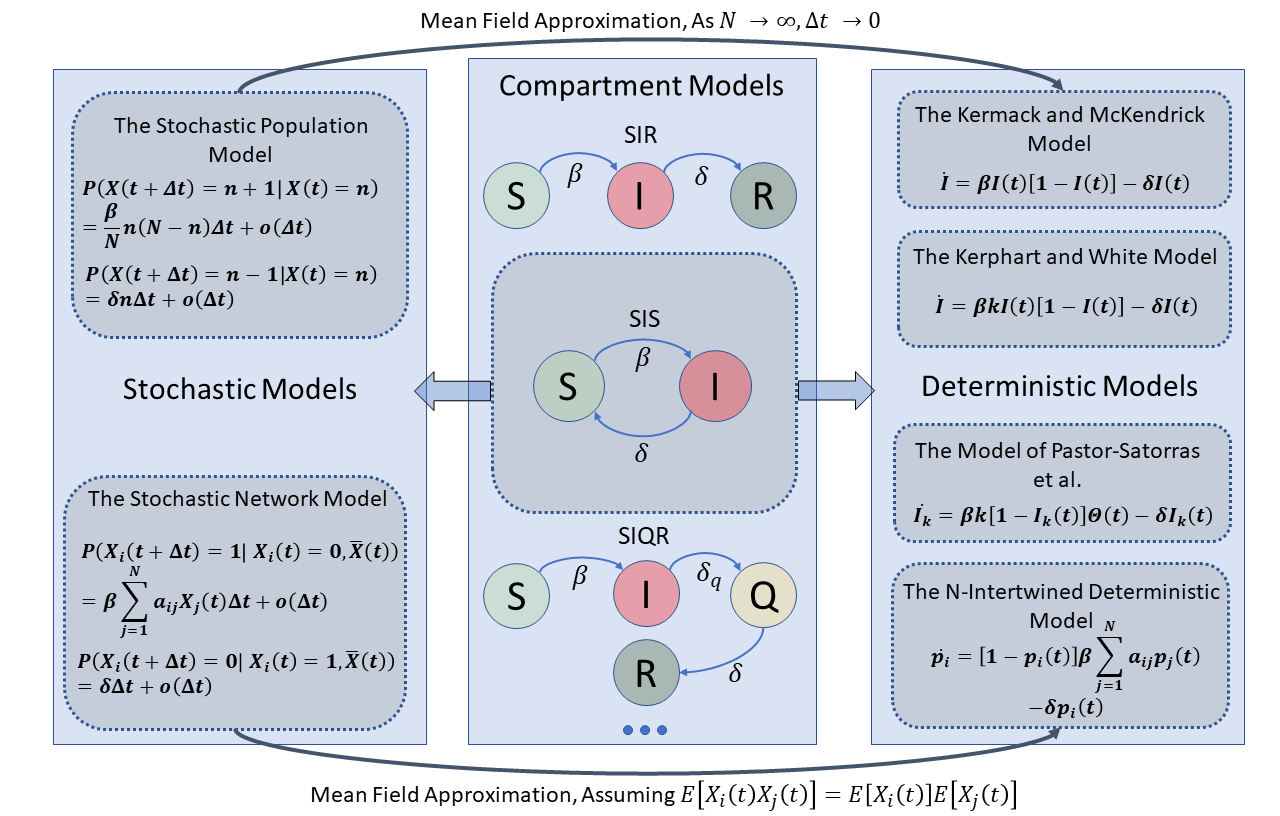}
\caption{A taxonomic summary of well-known epidemic models and their connection.  }
\label{fig:EpidemicModels}       
\end{figure*}

\subsection{Epidemic Models}\label{subsec:EpidemicModels}

The spreading of infectious diseases has affected human civilization since the age of nomadic hunter-gatherers. Mathematical models of epidemic spreading started to be proposed and studied since the beginning of the industrial revolution, with one of the earliest attempts to model infectious disease transmission mathematically by Daniel Bernoulli \cite{bernoulli1760essai}. Over the last $200$ years, many mathematical models of epidemics have been proposed and analyzed \cite{andersson2012stochastic,pare2020modeling,draief2009epidemics,van2008virus,pastor2001epidemic}. 

\paragraph{Compartment Models:} The basic models of many mathematical modeling of epidemics are the well-known compartment models \cite{kermack1932contributions}. In compartment models, every subject, based on the status, belongs to some compartment of the population at any given time. Common compartments include the susceptible (S), the exposed (E), the infected (I), the asymptomatic (A), and the recovered {who are immune to the disease (R).} Depending on the mechanisms of infectious diseases, different compartment epidemic models such as SIS (Susceptible-Infected-Recovered), SIR (Susceptible-Infected-Recovered), and SEIR (Susceptible-Exposed-Infected-Recovered) are studied and analyzed. Some infections, such as the common cold and influenza, do not confer any long-lasting immunity. Such infections can be modeled by the SIS model since individuals can become susceptible again after their recovery. If individuals recover with permanent immunity, the model is an SIR model. The SIR model has been used to study infectious diseases such as measles, mumps, and smallpox \cite{kermack1932contributions}. Many variants of compartment models have been proposed and studied in the past decades. For example, Rothe et al. \cite{rothe2020transmission} have looked into an SAIRS (susceptible-asymptomatic-infected-recovered-susceptible) to model the COVID-19 pandemic; Erdem et al. \cite{erdem2017mathematical} have investigated an SIQR (susceptible-asymptomatic-infected-quarantine-recovered) to study the effect of imperfect quarantine on the spreading of an influenza epidemic; Huang et al. \cite{huang2016novel} have proposed a model that connects the SIS and the SIR models, in which once recovered from an infection, individuals become less susceptible to the disease.

\subsubsection{Stochastic versus Deterministic Models}
To capture the dynamics of the epidemic-spreading process, we need a dynamic model to describe the evolution of the population in each compartment. In general, epidemic models are categorized into two groups: deterministic epidemic models and stochastic epidemic models, depending on the mathematical formulation. Deterministic models, oftentimes represented by a collection of ordinary differential equations (ODEs), have perhaps received more attention in the literature \cite{kermack1932contributions,pare2020modeling,pastor2001epidemic,van2008virus}. Their popularity is because deterministic models can become more complex yet still feasible to analyze, at least when numerical results are sufficient. In contrast, stochastic models, usually represented by Markov processes,  need to be fairly simple to be mathematically manageable\cite{andersson2012stochastic,draief2009epidemics}. 

There are, however, several advantages of stochastic epidemic models over deterministic epidemic models when the analysis is tractable. First, the epidemic spreading processes are stochastic by nature. For example, the disease transmission between individuals is more spontaneously described by probabilities than deterministic rules that govern the transmission. Second, stochastic modeling has its deterministic counterpart through mean-field analysis. 
The stochastic modeling provides a microscopic description of the epidemic process, while the deterministic counterpart is useful to describe the spreading of epidemics at a macroscopic level, e.g., the fraction of the infected population at a given time. When the number of individual $N$ is small, or the size of the infection is small in a large community,  the mean-field approximation will experience a considerable approximation error, and hence deterministic models may fail to accurately describe the spreading process \cite{van2008virus}. Third, deterministic models are incapable of capturing higher-order characteristics of the spreading process, such as variances, which are useful for the understanding of the uncertainties in the estimates. 

Overall, deterministic models and stochastic ones are complementary to each other. Deterministic models describe the spreading process at a macroscopic level and are more manageable mathematically, yet subject to assumptions on the spreading processes. Stochastic models approach the spreading processes from a microscopic point of view and offer a detailed description of the spreading process. In the following subsections, we introduce both deterministic and stochastic models using the SIS epidemic models as examples.

\subsubsection{Deterministic Models}\label{subsubsec:DeterministicModels}

Over the last century, increasingly sophisticated deterministic epidemic models have been proposed to capture the spreading processes on growingly complex and realistic networks.

In an SIS model, each individual in the system is either infected or susceptible. An infected node can infect its susceptible neighbors with an infection rate $\beta$. An infected individual recovers at recovery rate $\delta$. Once recovered, the individual is again prone to the disease. The simplest deterministic SIS model is introduced by Kermack and McKendrick \cite{kermack1932contributions}:
\begin{equation}\label{KMSISModel}
\begin{aligned}
    \dot{S} &= -\beta S(t) I(t) + \delta I(t),\\
    \dot{I} &= \beta S(t) I(t) - \delta I(t),
\end{aligned}
\end{equation}
where $S(t)$ is the fraction of the population who are susceptible, $I(t)$ is the fraction of the infected. In {(\ref{KMSISModel})}, the rate at which the fraction of infected individuals evolve is determined by the rate at which the infected population is recovered, i.e., $\delta I(t)$ and the rate at which the fraction of infected population grows, i.e., $\beta S(t) I(t)$. The latter rate captures the encounter between the fraction of susceptible individuals and the fraction of infected individuals. Simple deterministic models like (\ref{KMSISModel}) assume that the individuals in the population are homogeneously mixed; i.e., each individual is equally likely to encounter every other node. Such models have ignored the structure of the underlying network.

Starting from the 90s, new deterministic models have been proposed and studied to accommodate epidemic processes over more complex and realistic network structures. Kerphart and White \cite{kephart1992directed} investigated a regular graph with $N$ individuals where each individual has degree $k$. The Kerphart and White model is described by the ODE:
\begin{equation}\label{KWSISModel}
    \dot{I} = \beta k I(t)[1-I(t)] - \delta I(t),
\end{equation}
where the rate of infection is $\beta k I(t)(1-I(t))$ which is proportional to the fraction of susceptible individuals, i.e., $1-I(t)$. For each susceptible individual, the rate of infection is the product of the infection rate $\beta$ and the number of infected neighbors $kI(t)$. The Kermack and McKendrick model (\ref{KMSISModel}) and the Kerphart and White model (\ref{KWSISModel}) are referred to as ``homogeneous'' models since they assume that the underlying network has homogeneous degree distributions, i.e., each node in the network has the same degree.

With the emerging occurrence of complex networks in many social, biological, and communication systems, it is of great interest to investigate the effect of
their features on epidemic and disease spreading. Pastor-Satorras et al. \cite{pastor2001epidemic,pastor2001epidemic1} studied the spreading of epidemics on scale-free (SF) networks. In SF networks, the probability that an individual has degree $k$ follows a scale-free distribution $P(k)\sim k^{-\gamma}$, with $\gamma$ ranges from $2$ to $3$. It has been shown that many social networks such as collaboration networks, and computer networks such as the Internet and the World Wide Web exhibit such structure properties. The Pastor-Satorras model further divides individuals into sub-compartments based on their degrees, with $I_k(t)$ representing the proportion of infected individuals with a given degree $k$:
\begin{equation}\label{PSSISModel}
\dot{I}_k = \beta k [1-I_k(t)]\Theta(t) - \delta I_k(t),
\end{equation}
where 
$$
\Theta(t) = \sum_{k} \frac{k P(k)I_k(t)}{\sum_{k'} k' P(k')}
$$
represents the probability that any given link points to an infected individual. The rate at which the infected population grows is proportional to the infection rate $\beta$, the number of connections $k$, and the probability $\Theta(t)$ of linking to an infected individual. The model of Pastor-Satorras et al. incorporates the degree distribution of the underlying network and approximates the spreading processes on more complex and realistic networks.

To incorporate an arbitrary network characterized by the adjacency matrix $A$, Wang et al. \cite{wang2003epidemic} have proposed a discrete-time model that generalizes the Kerphart and White model (\ref{KWSISModel}) and the model of Pastor-Satorras et al. (\ref{PSSISModel}). Mieghem et al. \cite{van2008virus} have studied a continuous-time SIS model, called the $N$-intertwined deterministic model that generalizes the results in \cite{wang2003epidemic}. The $N$-intertwined deterministic model describes the spreading processes by
\begin{equation}\label{MieghemSIS}
\dot{p}_i = [1-p_i(t)]\beta \sum_{j=1}^N a_{ij} p_{j}(t) - \delta p_i(t),\ \ \ i=1,2,\cdots,N.
\end{equation}
where $p_i$ denotes the probability of individual $i$ being infected \cite{van2008virus} or the proportion of infected individuals in the sub-population $i$ \cite{fall2007epidemiological}. The element $a_{ij}$ in the adjacency matrix represents the connectivity between individual $i$ and individual $j$. The $N$-intertwined deterministic model provides a microscopic description of the spreading processes by incorporating the adjacency matrix that fully characterizes the underlying network. Other variants of the $N$-intertwined deterministic model (\ref{MieghemSIS}) have been recently studied in \cite{khanafer2014stability,pare2017epidemic}. To capture the heterogeneity of individuals' demographic or health situation, people consider a $N$-intertwined model with heterogeneous parameters:
\begin{equation}\label{HeteroMieghemSIS}
\dot{p}_i = [1-p_i(t)]\beta_i \sum_{j=1}^N a_{ij} p_{j}(t) - \delta_i p_i(t),\ \ \ i=1,2,\cdots,N,
\end{equation}
where $\beta_i$ describes the infection rate of individual $i$ and $\delta_i$ represents the recovery rate of individual $i$. Khanafer et al. have considered an $N$-intertwined mode in a directed network \cite{khanafer2014stability}. Par{\'e} et al. have extended the $N$-intertwined deterministic model to accommodate time-varying networks \cite{pare2017epidemic}.

Most studies of deterministic epidemic models are interested in analyzing the system equilibria (the limit behavior of the spreading process as time reaches infinity), characterizing the threshold that determines which equilibrium the system converges to, and evaluating stability properties of different equilibria \cite{van2008virus,pare2020modeling,pastor2001epidemic1,kephart1992directed,wang2003epidemic}. Define the effective spreading rate $\tau \coloneqq \beta /\delta$. One of the primary goals in most studies of deterministic epidemic models is to characterize the threshold $\tau_c$.  {If $\tau>\tau_c$, the epidemic persists and at least a nonzero proportion of the individuals are infected. If $\tau\leq \tau_c$, the epidemic dies out.} The Kerphart and White model (\ref{KWSISModel}) gives a ``steady-state'' epidemic threshold $\tau_{c,\textrm{KW}} = 1/k$, which is inversely proportional to the individuals' degree in a regular graph \cite{kephart1992directed}. The model of Pastor-Satorras et al. (\ref{PSSISModel}) provides a threshold $\tau_{c,\textrm{PS}}= {\langle k \rangle}/{\langle k^2 \rangle }$, with $\langle k \rangle = \sum_{k'} k'P(k')$ and $\langle k^2\rangle = \sum_{k'} {k'}^2 P(k')$. {Hence, in SF networks, there will be an absence of threshold if  $\langle k^2 \rangle  \rightarrow \infty$, meaning even a disease with a low infection rate can cause an outbreak in such a network.} The epidemic threshold for the $N$-intertwined deterministic model (\ref{MieghemSIS}) is specified by $\tau_{c,\textrm{VPM}} = 1/ \lambda_{max}(A)$, where $\lambda_{max}(A)$ the special radius of the adjacency matrix $A$ \cite{van2008virus}. For the heterogeneous $N$-intertwined deterministic model (\ref{HeteroMieghemSIS}), the infectious disease will die out when $\lambda_{max}(BA-D)\leq 0$. Otherwise, an outbreak will occur. Here, $D=\textrm{diag}(\delta_1,\delta_2,\cdots,\delta_N)$, $B=\textrm{diag}(\beta_1,\beta_2,\cdots,\beta_N)$, and $A$ is the adjacency matrix of the underlying network. For a comprehensive review of deterministic epidemic models, their {``outbreak''} thresholds and stability properties, one can refer to \cite{nowzari2016analysis,pare2020modeling}.

\subsubsection{Stochastic Models}\label{subsubsec:StochasticModels}

In this subsection, we present several well-known stochastic epidemic models based on SIS compartment models. {The counterparts for the SIR type of epidemics can be formulated by adding a new compartment R and describing the transition rule between compartments S, I, and R}. First, we consider stochastic epidemic models without network structures, which we refer to as stochastic population models.

Consider a population of $N$ individuals. Recall that the rate of infected individuals infecting someone else is denoted by $\beta$. Each infected individual recovers and becomes susceptible at rate $\delta$. Assume that individuals encounter each other uniformly at random from the whole population. Let $x(t)$ be the number of infected individuals at time $t$. Then, $\{X(t)\}_{t\geq 0}$ is a Markov jump process with state space $\mathbb{N}$, with transition rates $\beta n(1-n/N)$ from state $n$ to state $n+1$, and $\delta n$ from state $n$ to state $n-1$. The transition probability hence is given as 
\begin{equation}\label{StochasticPopModel}
\begin{aligned}
    &\mathbb{P}(X(t+\Delta t)=n+1 \vert\ X(t) =n) = \frac{\beta}{N}n(N-n) \Delta t +  o(\Delta t),\\
    &\mathbb{P}(X(t+\Delta t) = n-1 \vert X(t) =n ) = \delta n\Delta t + o(\Delta t),\\
    &\mathbb{P}(X(t+\Delta t) = n \vert X(t) =n ) = 1 - \Delta t \left(\frac{\beta}{N}n(N-n) + \delta n \right) + o (\Delta t).
\end{aligned}
\end{equation}
The stochastic population model (\ref{StochasticPopModel}) does not consider network structures and assumes that each individual encounters each other uniformly. It is clear that the stochastic population model (\ref{StochasticPopModel}) is a Markov jump process with an absorbing state $X = 0$, i.e., the state where no individual is infected. When the process enters the absorbing state, the infection dies out. 

{Indeed, stochastic epidemic models with a finite population admit an absorbing state at the origin (i.e., no infection) and a degenerate stationary distribution that has a probability $1$ at the origin. The stochastic model gives a prediction that extinction will ultimately occur for any given initial distribution, regardless of the value $\beta/\delta$. However, the time to extinction does change drastically with $\beta/\delta$. When $\beta/\delta$ is distinctly less than $1$, the ``die-out'' time is small and almost independent of $\beta/\delta$. However, the ``die-out'' time increases exponentially in the number of population $N$ if $\beta/\delta$ is distinctly larger than $1$ \cite{kryscio1989extinction}. The ``die-out'' time can last longer than the entire human history for a small population. Hence, the distribution of the number of infected individuals during the long waiting time before extinction is close to the distribution of the same random variable under the condition that extinction has not occurred \cite{naasell1996quasi}. This is called the quasi-stationary distribution of the stochastic model. Analyzing the quasi-stationary distribution for various stochastic models becomes the focus of many papers \cite{naasell1996quasi,kryscio1989extinction,allen2008introduction}}. 

Another common result in the studies of stochastic epidemic models dwells is to determine
the mean-field deterministic model associated with the stochastic model when the population becomes large. Indeed, as $N$ goes to infinity, the trajectory $t\rightarrow X(t)/N$ converges to the solution $I(t)$ of the Kermack and McKendrick deterministic model (\ref{KMSISModel}), i.e., $\lim_{n\rightarrow \infty } \sup_{0\leq t \leq T} \vert X(t)/N - I(t) \vert = 0$ almost surely. The proof of the convergence is underpinned by the well-known Kurtz's theorem. One can refer to Section 5.3 of \cite{draief2009epidemics} for more details.

We now introduce the stochastic network models that incorporate network structures. Define $X_i(t)$ as the epidemic state of individual $i$. The individual is infected if $x_i(t)$, and healthy (susceptible) if $X_i(t)=0$. Infected individuals recover at rate $\delta$, while susceptible individuals become infected at rate $\beta \sum_{j=1}^N a_{ij}X_j$, i.e., the product of the the infection rate and the number of infected neighbors. Let $\bar{X}(t) = (X_1(t),\dots, X_N(t))^T$ be the entire epidemic state of the whole population. The SIS stochastic network model can be expressed by the following Markov process: 
\begin{equation}\label{2NExactSISModel}
\begin{aligned}
    &\mathbb{P}(X_i(t+\Delta t) =0\vert\ X_i(t)=1, \bar{X}(t)) = \delta \Delta t + o(\Delta t),\\
    &\mathbb{P}(X_i(t+\Delta t) = 1\vert\ X_i(t)=0,\bar{X}(t)) = \beta \sum_{j=1}^N a_{ij} X_j(t) \Delta t +  o(\Delta t),
\end{aligned}
\end{equation}
for $i=1,2,\cdots,N$. The stochastic network model (\ref{2NExactSISModel}) has $2^N$ states among which there is an absorbing state (i.e., a state where no individual is infected) reachable from any state $\bar{X}(t)$ with non-zero probability. Hence, this model indicates that the epidemic will die out almost surely in finite time. A more meaningful way to describe the spreading process is through probabilistic quantities. Indeed, the probability that the epidemic will not die out at time $t$ is bounded as follows \cite{draief2009epidemics}
\begin{equation}\label{DieOutProbability}
\mathbb{P}(\bar{X}(t) \neq \mathbf{0}) \leq \exp{(\left(\beta \lambda_{max}(A)-\delta)t\right)} \sqrt{N \sum_{i=1}^N X_i(0)},
\end{equation}
where $A$ is the adjacency matrix of the underlying network, $\lambda_{max}(A)$ is its spectral radius. From (\ref{DieOutProbability}), how fast the epidemic will die out depends on the effective spreading rate $\tau \coloneqq \beta /\delta$ and the spectral radius of the underlying network. Let $T$ denote the time to absorption state (the time when the epidemic dies out). An application of (\ref{DieOutProbability}) leads to the results regarding the expected extinction of the epidemic \cite{draief2009epidemics}: Given arbitrary initial condition $\bar{X}(0)\in\{0,1\}^N$, if $\tau< 1/(\lambda_{max}(A))$,
\begin{equation}\label{ExpectedDieOutTime}
\mathbb{E}[T] \leq \frac{\log N + 1}{\delta - \beta \lambda_{max}(A)}.
\end{equation}
If $\tau \geq 1/(\lambda_{max}(A))$, the expected extinction time increases exponentially as the number of individuals $N$ increases, i.e., $\mathbb{E}[T] = O(cN)$, where $c$ depends on the effective spreading rate $\tau$ and the network structure \cite{ganesh2005effect}. Loosely speaking, the spectral radius $\lambda_{max}(A)$ quantifies ``how tightly the underlying network is connected''. The results (\ref{DieOutProbability}) and (\ref{ExpectedDieOutTime}) align well with the intuition that it is easier for an infectious disease to grow on a more tightly connected network. Letting $\Delta t$ goes to zero, the dynamics of $\mathbb{E}[X_i(t)]$ can be written as
\begin{equation}\label{ExpectationODE}
\dot{\mathbb{E}}[X_i(t)] = \mathbb{E}\left[ (1-X_i(t))\beta \sum_{j=1}^N a_{ij} X_j(t) \right] - \delta \mathbb{E}[X_i(t)],\ \ \ i=1,2,\cdots,N.
\end{equation}
Alleviating the complication of the term $\mathbb{E}[X_i(t)X_j(t)]$ by assuming $\mathbb{E}[X_i(t)X_j(t)] = \mathbb{E}[X_i(t)]\mathbb{E}[X_j(t)]$ for all $i\neq j$, one can recover from (\ref{ExpectationODE}) the $N$-intertwined deterministic model (\ref{MieghemSIS}), which re-states below:
$$
\dot{p}_i = [1-p_i(t)]\beta \sum_{j=1}^N a_{ij} p_{j}(t) - \delta p_i(t),\ \ \ i=1,2,\cdots,N.
$$
Since $X_i(t)\in \{0,1\}$, $p_i(t) \coloneqq \mathbb{E}[X_i(t)] = \mathbb{P}(X_i(t)=1)$ represents the probability that individual $i$ is infected at time $t$. Since $\mathbb{E}[X_i(t)X_j(t)] = \mathbb{E}[X_i(t)]\mathbb{E}[X_j(t)]$ is not necessarily true, the $N$-intertwined deterministic model (\ref{MieghemSIS}) serves as an approximation of the stochastic network model (\ref{2NExactSISModel}). Indeed, it is shown that the expected values $p_i$ in (\ref{MieghemSIS}) are upper bounds on the actual probabilities given by (\ref{2NExactSISModel}) \cite{van2008virus,cator2012second,nowzari2016analysis,pare2017epidemic}. Furthermore, \cite{van2015accuracy} investigates how accurate the deterministic approximations (\ref{MieghemSIS}) are in describing the
stochastic model (\ref{2NExactSISModel}).

So far, we have introduced main stochastic and deterministic epidemic models that are well studied in the literature. To offer an overview of these models and their connection, in {Figure~\ref{fig:EpidemicModels}}, we present an illustrative taxonomy of the epidemic models we introduced so far. Apart from models that are based on Markov processes and ODEs, there are epidemic models based on stochastic differential equations \cite{allen2007modeling,gray2011stochastic}, purely data-driven approaches \cite{chimmula2020time}, or spatial modeling \cite{rhee2011levy,valler2011epidemic,huang2016epidemic,possieri2019mathematical}. {Even though the focus of this review is on actual virus epidemics, there has been an abundant number of papers studying the spread of malware, Trojans, worms on mobile wireless networks \cite{khouzani2012maximum,altman2012epidemic,altman2010optimal}, social networks \cite{altman2013stochastic,masson2017posting,netrapalli2012learning}, Delay Tolerant Networks \cite{panda2014tracking,ali2012estimating}, general computer network \cite{legenvre2016potential}. These papers introduced many useful techniques and insightful results applied to analyzing models of actual epidemics \cite{khouzani2012maximum, legenvre2016potential, altman2012epidemic}. In this review, we will introduce several papers that do not necessarily focus on actual epidemics but provide useful insights for models of actual epidemics \cite{huang2020differential,kephart1992directed,khouzani2011saddle, trajanovski2015decentralized, trajanovski2017designing,zhao2018virus}. }

With a basic understanding of the epidemic spreading processes and their stochastic and deterministic modeling, the next section introduces how decisions can be made based on these models and how decision-making can in turn affect the spreading processes.

\subsection{Decision Models} \label{subsec:decisionmodels}

The epidemic models introduced in the previous subsection have described the spreading process when there is no human intervention. In practice, {once an epidemic starts to prevail, human will reacts accordingly by taking interventions.} The human intervention creates an considerable impact on the spreading processes. Hence, the modeling of infectious diseases needs to take it into consideration to provide a consolidated understanding of the epidemic spreading processes in the human population. 
To better understand epidemic spreading processes in human population, we introduce a holistic framework that incorporates epidemic models and decision models, illustrated in {Figure~ \ref{fig:EpidemicDecisionLoop}}.

\subsubsection{What Interventions to Take?}\label{subsec:interventions}

{In the real world, different entities play different roles in making decisions to combat the spread of the disease. Such entities can be individuals, households, and organizations. Decision making consists of the choices of individuals and the strategies of stakeholders at different hierarchical levels. Decision models considered in the existing literature mainly consider two types of decision-makers: a central planner \cite{xu2015competition,yang2021modeling,preciado2013optimal,kohler2020robust,ansumali2020modelling,parino2021Modelling,shen2014differential,pejo2020corona,dashtbali2020optimal,khouzani2011saddle,aurell2020optimal,pezzutto2021smart,hota2020closed,watkins2016optimal,ogura2016efficient,hota2016optimal,li2017minimizing,zhao2018virus,li2019suboptimal,watkins2019robust,zhang2019differential,di2020covid,huang2020differential,nowzari2016analysis,chen2021optimal,wang2020modeling,mai2018distributed,xue2018distributed}, which works for social benefits by conducting mechanism design and/or applying enforceable measures directly to the general public, and a collection of individuals with various types \cite{trajanovski2015decentralized,reluga2010game,theodorakopoulos2012selfish,hota2016interdependent,hota2019game,huang2020differential,aurell2020optimal,adiga2016delay,zhang2013braess,dashtbali2020optimal,eksin2019control,hayel2017epidemic,bauch2004vaccination,feng2016epidemic,feng2016epidemicPA,huang2019achieving,breban2007mean,trajanovski2017designing} who have their own goals. Considering only one central planner simplifies the case in the real world and allows modelers to focus on a particular aspect of decision making amid the epidemic and derive results that explain certain phenomena.}

Decisions makers can take either non-pharmaceutical interventions (NPIs) or pharmaceutical interventions or both. For a central planner, non-pharmaceutical interventions include but not limit to requiring mandatory social distancing, enforcing lockdown, quarantining infected individuals, and deploying protective resources such as masks, gloves, gowns, and testing kits. Pharmaceutical interventions are related with the availability of vaccines or antidotes. A central planner's decision may involve vaccine distributions, antidote allocations, treatment prioritization, etc. For individuals, possible non-pharmaceutical interventions are wearing a mask, practicing social distancing, self-quarantine, etc. Pharmaceutical interventions, such as getting vaccinated, seeking for treatment, securing an antidote,  are usually adopted by individuals to protect themselves in the epidemic.

\subsubsection{How are Interventions modeled?}\label{subsec:HowInterventionsModelled}

To understand the coupling between the decision models and the epidemic models, we need to figure out how non-pharmaceutical and pharmaceutical interventions can be modeled and incorporated into the epidemic models. Generally, non-pharmaceutical interventions help curb the spreading by either reducing the interaction between individuals (e.g., avoiding crowds, social distancing, lockdown, and quarantine) or utilizing protective resources (e.g., wearing a mask and  frequent use of hand sanitizer). 

In networked epidemic models such as the $N$-intertwined deterministic model (\ref{MieghemSIS}) and the stochastic network model (\ref{2NExactSISModel}), the interaction between individuals is usually captured by the network topology. In an unweighted network, the adjacency element $a_{ij}=1$ means that there exist interactions between individual $i$ and individual $j$. Otherwise, $a_{ij}=0$. Some existing {papers} use the adjacency elements $a_{ij}\in\{0,1\}$ to describe strict measures that completely cut down the interaction between individuals such as lockdown and quarantine \cite{watkins2019robust,xue2018distributed,eksin2017disease}. For example, if individual $i$ is quarantined, then $a_{ij}=0$ for all $j\in\mathcal{N}$. To capture the effect of {measures do not require complete isolation} such as social distancing, several {papers} have considered weighted networks to describe the intensity of interactions \cite{pezzutto2021smart,huang2020differential}. For example,  \cite{huang2020differential} uses a weight coefficient $w_{ij}\in[0,1]$ to describe the intensity of the interaction between individual $i$ and $j$. In epidemic models that do not capture the complete topology (e.g., the Kermack and McKendrick  model (\ref{KMSISModel}), the Kerphart and White model (\ref{KWSISModel}), the model of Pastor-Satorras et al. (\ref{PSSISModel}), and the stochastic population model (\ref{StochasticPopModel})),  the reduced interaction between individuals is modeled by a scaled infection rate $\alpha\beta$, where $\alpha\in[0,1]$ is the scaling factor and $\beta$ is the normal infection rate when there is no intervention  \cite{pejo2020corona,reluga2010game,kohler2020robust,ansumali2020modelling,parino2021Modelling,dashtbali2020optimal}. For example, in \cite{kohler2020robust}, $\alpha$ captures how well people practice social distancing. The better people practice social distancing, the smaller the factor $\alpha$ is. An alternative way of modeling interventions such as lockdown and quarantine is to create a new compartment \cite{erdem2017mathematical,li2019suboptimal,yang2021modeling}. In \cite{li2019suboptimal}, the central planner decides the number of infected individuals to be quarantined to curb the spreading. The authors introduce a new compartment called `Quarantine' to model infected individuals being selected for quarantine. Once individuals are quarantined, they will not infect susceptible individuals and will re-enter the Susceptible compartment after recovery.

Utilizing protective resources such as wearing a mask and using hand sanitizer helps individuals protect themselves from infection without reducing their interaction with other individuals. Such interventions are also captured by a scaled infection rate $\alpha \beta$ \cite{ogura2016efficient,nowzari2016analysis} with $\alpha \in [0,1]$, which describes the fact that when contacting infected individuals while wearing a mask, susceptible individuals will less likely to be infected. 

{Pharmaceutical interventions refer to the use of preventive medicines, vaccines, antidotes, or effective treatment methods when available. One example of preventive medicines is pre-exposure prophylaxis, which is proved effective in limiting HIV spread \cite{kim2014mathematical,pretorius2010evaluating}. Preventive medicines lower the infection rate $\beta$ of the target diseases. Individuals need frequent use of the medicine (e.g., daily or twice a day) to maintain a lower rate of infection to the disease. Preventive medicines protect individuals for a shorter period than vaccines \cite{pretorius2010evaluating}. Individuals can prolong the protection by constantly taking preventive medicines. Hence, to model using preventive medicines as an intervention, modelers use different ways to describe the cost and inconvenience of using preventive medicines than getting a vaccine.} Many researchers have studied the effect of vaccination on the spreading process \cite{gubar2015two,preciado2013optimal,andersson2012stochastic,hota2019game,trajanovski2015decentralized,adiga2016delay,bauch2004vaccination,li2017minimizing,mahrouf2020non}, which can be modeled in several ways. One way is to reduce the infection rate between an infectious and a vaccinated individual to $\beta_{vac}<\beta$ while those who are not vaccinated suffer a higher infection rate $\beta$ \cite{andersson2012stochastic,preciado2013optimal,hota2019game,adiga2016delay,bauch2004vaccination,li2017minimizing}. How small $\beta_{\textrm{vac}}$ is depends on the efficacy of the vaccine distributed.  Another way to create a new compartment called `Vaccinated' and the rate at which individuals exit this compartment captures the protection duration of the vaccines\cite{mahrouf2020non,abouelkheir2019optimal}. The use of antidotes and the deployment of mass treatment accelerate the recovery process. As a result, The use of antidotes and the deployment of mass treatment are usually modeled by a recovery rate $\bar{\delta}$ higher than the natural recovery rate $\delta$ \cite{chen2021optimal}.

\subsubsection{When are Interventions Taken?}\label{sebsec:whenInterventions}

{Timing} plays a significant role in the coupling between the epidemic spreading processes and the decision models. Depending on when decisions are made, decision-making can be categories into pre-epidemic decision-making \cite{hossain2020explainable,herrera2016disease,sparks2011optimal,hota2016optimal,nowzari2016analysis,preciado2013optimal,blume2013network,trajanovski2017designing}, during-epidemic decision-making \cite{mai2018distributed,pejo2020corona,saha2014equilibria,bauch2004vaccination,hota2016interdependent,hota2019game,li2017minimizing,xue2018distributed,chen2021optimal,hota2020impacts,shen2014differential,eksin2019control,akhil2019mean,dashtbali2020optimal,eksin2017disease,reluga2010game,khouzani2011saddle,adiga2016delay,huang2020differential,di2020covid,farhadi2019efficient,hota2020closed,pezzutto2021smart,watkins2019robust,theodorakopoulos2012selfish,nowzari2016analysis,li2019suboptimal,breban2007mean,liu2012impact,zhang2013braess,trajanovski2015decentralized}, and post-epidemic decision-making\cite{xue2018distributed,huang2020differential,patro2020towards,bieck2020redirecting}. Pre-epidemic decision-making refers to the decisions made before an epidemic happens or at the beginning of the pandemic. Most papers focus on network design/formation problems in which a virus resistant network with guaranteed performance is designed/formed \cite{hota2016optimal,nowzari2016analysis,blume2013network,trajanovski2017designing}. Some researchers study the optimal design of an epidemic surveillance system on complex networks that helps detect an epidemic at an early stage \cite{hossain2020explainable,herrera2016disease,sparks2011optimal}. Early epidemic detection allows a central planner to kill the spreading at its infancy. Post-epidemic decision-making happens at the very end of an epidemic or after an epidemic dies out. The post-epidemic decision-making addresses problems such as how to safely lift restrictions, how to reverse the interventions taken during the epidemic season \cite{xue2018distributed,huang2020differential,patro2020towards,bieck2020redirecting}.

This review focuses on during-epidemic decision-making problems where decisions are made while an epidemic is present. The dynamics of the epidemic spreading processes and the dynamics of the decision adaptation may evolve at different time scales. Some interventions such as getting vaccinated (if one obtains life-long protection from the vaccine), taking an antidote, and distributing curing resources are irreversible \cite{mai2018distributed,bauch2004vaccination,hota2016interdependent,hota2019game,trajanovski2015decentralized}. For such interventions, decisions were only made once. In other studies, authors assume that decisions are made once and for all and remain fixed over the spreading process or that decisions are made at a much smaller frequency than the epidemic spreading process \cite{pejo2020corona,saha2014equilibria,chen2021optimal}. These assumptions and the irreversibility of some interventions make it possible to formulate a static optimization \cite{mai2018distributed,chen2021optimal} or game \cite{bauch2004vaccination,hota2016interdependent,hota2019game,trajanovski2015decentralized,pejo2020corona,saha2014equilibria} problem to study the decision-making during the epidemic spreading. The objective functions of these static optimizations or game problems only involve the limiting behavior of the epidemic models \eqref{KMSISModel}-\eqref{2NExactSISModel}, e.g., the infection level at equilibria (when the epidemic model reaches a steady state) \cite{trajanovski2015decentralized,chen2021optimal,bauch2004vaccination,saha2014equilibria}, whether the threshold condition is met (i.e., whether the epidemic will eventually die out) \cite{nowzari2016analysis}.


If the epidemic spreading processes and the decision models evolve at the same time scale, adaptive strategies are employed in which the decisions are adapted at the same pace as the epidemic propagates. For example, a central planner distributes testing kits based on the daily infection data. This creates a real-time feedback loop in the human-in-the-loop epidemic framework shown in {Figure~\ref{fig:EpidemicDecisionLoop}}. In this case, individuals take interventions such as whether to wear a mask, conduct social distancing, or stay self-quarantined based on currently perceived information; the central planner adapts his/her interventions according to the observed infection status of the whole population in real-time. The strategies of individuals and the central planner is represented by a map that maps the information they received so far to an action that describes the interventions being taken. Depending on the choice of epidemic models, researchers employ different tools to study the human-in-the-loop epidemic framework depicted in {Figure~\ref{fig:EpidemicDecisionLoop}}. When deterministic models are employed to describe the epidemic spreading processes, optimal control \cite{dashtbali2020optimal,di2020covid,hota2020closed,watkins2019robust,nowzari2016analysis,li2019suboptimal}, differential game \cite{huang2020differential,shen2014differential,dashtbali2020optimal,reluga2010game,khouzani2011saddle}, or evolutionary game theory \cite{theodorakopoulos2012selfish} have been applied to study the human-in-the-loop epidemic framework. When stochastic models are used, the human-in-the-loop epidemic framework is often  modeled by Markov decision processes \cite{pezzutto2021smart,yaesoubi2011generalized,gast2012mean} and stochastic games \cite{hota2020impacts,eksin2019control,akhil2019mean,eksin2017disease}. Some research papers consider seasonal epidemics and at each epidemic season, decisions are made once. People adapt their decisions based on the payoff of the prior seasonal epidemic. This type of decision-making problem under seasonal epidemics is solved by analyzing repeated games \cite{breban2007mean,liu2012impact} or evolutionary games \cite{liu2012impact}.

\subsubsection{Who are the decision-makers?}\label{subsec:decisionMakers}

For the human-in-the-loop epidemic framework, decision-makers usually involve a central planner that represents the central authority and individuals that represent the general public \cite{huang2020differential}. The central planner can be an effective central government when fighting against an epidemic, or a network operator whose users are obliged to abide by the company security policy. An individual can be a individual citizen of a society \cite{van2008virus}, a local community \cite{fall2007epidemiological}, {public authorities of different countries \cite{maggi2014coordination}} or a user in a computer network \cite{zhao2018virus}.

A central planner cares about the welfare of the whole population such as the number of infected individuals in the entire population, the well-being of the economy \cite{xu2015competition,yang2021modeling,preciado2013optimal,kohler2020robust,ansumali2020modelling,parino2021Modelling,shen2014differential,pejo2020corona,dashtbali2020optimal,khouzani2011saddle,aurell2020optimal,pezzutto2021smart,hota2020closed,watkins2016optimal,ogura2016efficient,hota2016optimal,li2017minimizing,zhao2018virus,li2019suboptimal,watkins2019robust,zhang2019differential,di2020covid,huang2020differential,nowzari2016analysis,chen2021optimal,wang2020modeling,mai2018distributed,xue2018distributed}. Individuals concern about their own interests, which include his/her own infection risk, the inconvenience of wearing a mask, and the monetary cost of getting a effective treatment \cite{trajanovski2015decentralized,reluga2010game,theodorakopoulos2012selfish,hota2016interdependent,hota2019game,huang2020differential,aurell2020optimal,adiga2016delay,zhang2013braess,dashtbali2020optimal,eksin2019control,hayel2017epidemic,bauch2004vaccination,feng2016epidemic,feng2016epidemicPA,huang2019achieving,breban2007mean,trajanovski2017designing}. Due to the selfishness of the individuals, the goal of an individual is not well aligned and sometimes conflicts with the goal of a central planner \cite{huang2020differential, eksin2017disease}. For example, the infected individuals, with no infection risk anymore, might be reluctant to take preemptive measures to avoid spreading the disease \cite{eksin2017disease}. It is shown that there will be an increase in the number of infected individuals if they optimize for their own benefits instead of complying with the rules applied by the central planner \cite{huang2020differential}. Indeed, enforcing a strict protocol can be costly and sometimes impossible for the central planner. As a result, central authorities should ask themselves whether they can offer the public sufficient incentives that are acceptable by the individuals and sufficiently strong to combat the epidemic. A recent example is that, to reach herd immunity, the Ohio state of the United States will give $5$ people $1$ million each in COVID-19 vaccine lottery to combat the hesitancy of getting a COVID-19 vaccine \cite{dareh2021vaccinated}. Hence, instead of solving an optimization problem or a game problem directly, some papers have looked into the mechanism design problem or the information design problem \cite{zhang2021informational}, on behalf of the central planner, that incorporates both the global state of the whole population and the individual's choice into designing incentives to combat the epidemic \cite{aurell2020optimal,huang2020differential,farhadi2019efficient,pejo2020corona,breban2007mean,li2017minimizing,omic2009protecting}. 

\subsubsection{Information Matters}

{Decision-makers} rely on what information they have to make decisions. The information available to decision-makers at the time when the decision is made plays a crucial role in the human-in-the-loop epidemic framework \cite{di2020covid}. For example, the severity of COVID-19 infection, perception of government responses, media coverage, acceptance of COVID-19-related conspiracy theories lead to a change of people's attitude about wearing masks during {the COVID-19 pandemic} \cite{rieger2020german,romer2020conspiracy,Fisher2020}. Many studies assume that perfect information, including the health status of every individual and complete knowledge of the network topology, is available to decision-makers at all times \cite{eksin2019control,shen2014differential,dashtbali2020optimal,farhadi2019efficient,eksin2017disease,hota2019game,adiga2016delay,nowzari2016analysis,li2019suboptimal}.  However, in epidemics, acquiring perfect, accurate, and timely information regarding the spreading process is arduous if not impossible \cite{bhattacharyya2010game}. For example, obtaining an estimate of the number of infected individuals requires testing at scale, which can be challenging to implement in rural areas \cite{mercer2021testing}. Also, testing results can be delayed due to a high testing demand and a long sample analysis time. Hence, some researchers investigated the decision-making based on an estimated disease prevalence from available data \cite{hota2020impacts,pezzutto2021smart,watkins2019robust} and the effect of delayed information in decision-making \cite{zhu2019stability}. {In game theory, a specific type of strategic games that deal with incomplete information are called Bayesian games. A few papers have leveraged the concept of Bayesian games to deal with the incomplete information when only part of the information or some statistics about the information are revealed to the players \cite{Grottke2016,tsemogne2021game}.}

Individuals can receive information from mass media (global broadcasters) such as TV, radio, newspaper, and official accounts on social media and/or from local contacts such as friends, family members, and connections on social media. During an epidemic, individuals may receive two levels of information: one is statistical information that describes the overall prevalence of the epidemic such the number of positive cases, the number of hospitalized patients, and the death toll; the other is local information such as whether people with close connections are infected or not, risk level in one's neighborhood \cite{lagos2020games,granell2014competing}. \cite{wang2020epidemic,granell2014competing,funk2009spread} investigate how the word-of-mouth type of information spreading affects individuals' behavior and hence, alters the spreading processes.

Individuals may suffer from inaccurate information from unreliable resources. An example would be information obtained from social media or by word of mouth in a spatially or culturally isolated community or neighborhood. The perceived information of an individual may not necessarily reflect the actual prevalence of the infectious disease. Such incomplete or biased information about the epidemic together with strong prior beliefs may impede individuals from taking rational and reasonable responses to protect themselves and others \cite{reluga2010game}. Information released by central authorities also plays a significant role in individuals' decision-making. Responsible central authorities should therefore not only fight against the epidemic-related misinformation but also conduct information design to curb epidemic spreading and even panic spreading.

\section{Game-Theoretic decision-making in Epidemics}

Game theory, in a nutshell, is a powerful mathematical tool of modeling how people make strategic decisions within a group \cite{fudenberg1991game,bacsar1998dynamic}. In the last decade, there has been a surge in research studies in game-theoretic decision-making amid an epidemic \cite{theodorakopoulos2012selfish,hota2016interdependent,hota2016optimal,li2017minimizing,hota2019game,zhang2019differential,huang2020differential,khouzani2011saddle,aurell2020optimal,adiga2016delay,zhang2013braess,dashtbali2020optimal,eksin2017disease,eksin2019control,pejo2020corona,hayel2017epidemic,bauch2004vaccination,shen2014differential,huang2019achieving,breban2007mean,reluga2010game,trajanovski2015decentralized,xu2015competition,saha2014equilibria,hota2020impacts,farhadi2019efficient,liu2012impact,trajanovski2017designing,omic2009protecting,bhattacharyya2010game,lagos2020games,chang2020game,amaral2021epidemiological,ibuka2014free,reluga2006evolving}. The reason behind the surge is three-fold. 

First, centralized decision-making becomes less practical for large-scale networked systems such as human contact networks and most computer networks. Computing centralized protection strategies faces the challenge of scalability when they are applied to very large networks \cite{mai2018distributed,xue2018distributed,hota2016interdependent}. Also, it requires a high level of information granularity for a central authority to make satisfactory centralized decisions for most individuals. The central authority has to gather a huge amount of local information, which  not only is challenging to implement but also creates privacy issues and management overheads. In contrast, decentralized decision-making is more reliable and practical since local entities decide their own protection strategies satisfying high-level guidelines provided by the central authority. Second, Self-interested individuals in the
midst of the epidemic might not be willing to comply with the suggested protocols \cite{aurell2020optimal,bauch2004vaccination}. This is because, as we have explained in Section \ref{subsec:decisionMakers}, there is a misalignment of individual interests. Individuals concern less about and societal interests that are major concerns of the central authority. There also exists a misalignment of interests between individuals and interdependencies among the individuals. Each individual has choices, but the payoff for each choice depends on choices made by others. 
Third, game theory, as a mature and broad field, provides a plethora of useful solution concepts and analytical techniques that can model and explain human decision-making. For example, the self-interested strategy maximizing individual payoff is called the Nash equilibrium in game theory \cite{bacsar1998dynamic}. Through a Stackelberg game framework, a central authority can design incentives for the public individuals to combat the epidemic \cite{aurell2020optimal}. Many infectious disease models usually do not incorporate  human behaviors that change as the epidemic evolves and the information spreads over the network. Dynamic game theory, which has been applied in many dynamic settings such as management science \cite{bagagiolo2014mean}, labor economics \cite{liu2020stochastic}, and cybersecurity \cite{huang2020dynamic}, delivers a powerful paradigm to capture dynamic human behaviors \cite{bauch2004vaccination}. We start the introduction of game-theoretic models in epidemics by presenting a taxonomy in the next section.

\subsection{A Multi-Dimensional Taxonomy of Game-Theoretic Models in Epidemics}

The synthesis of game-theoretic models and epidemic models roots in the coupling between decision models and epidemic models. The choice of game-theoretic models depends on multiple factors such as who the decision-makers are, what interventions decision-makers can take, what information decision-makers know and etc. {Existing literature mainly studied the following five types of games: static games \cite{hota2018game,hota2019game,hota2016optimal,huang2019achieving,omic2009protecting,pejo2020corona,saha2014equilibria,trajanovski2015decentralized,trajanovski2017designing,xu2015competition}, discrete-time stochastic games \cite{eksin2019control,eksin2017disease,lagos2020games}, differential games \cite{reluga2010game,aurell2020optimal,dashtbali2020optimal,huang2019achieving,huang2020differential,khouzani2011saddle,shen2014differential}, repeated games \cite{adiga2016delay,breban2007mean,li2017minimizing,huang2019game}, and evolutionary games \cite{hayel2017epidemic,amaral2021epidemiological,zhang2013braess,reluga2006evolving,poletti2009spontaneous}.}

In static game frameworks, researchers have incorporated the epidemic models by only considering the limiting behavior of these models \cite{hota2019game,hota2016optimal,omic2009protecting,saha2014equilibria,trajanovski2015decentralized,trajanovski2017designing,xu2015competition}. Here, we use \cite{omic2009protecting} as an example. In \cite{omic2009protecting}, J. Omic et al. captures the risk of infection using the limiting behavior of the heterogeneous $N$-intertwined deterministic model (\ref{HeteroMieghemSIS}). Let $p_{i\infty} $ be the steady state of model ({\ref{HeteroMieghemSIS}}) for each individual $i\in\mathcal{N}$. Letting $\dot{p}_i =0$, one obtains
\begin{equation}\label{eq:steadystateintertwined}
p_{i\infty} = \frac{ \sum_{j=1}^N a_{ij}p_{j\infty}}{\beta_i\sum_{j=1}^N a_{ij}P_{j\infty} + \delta_i},\ \ \ \textrm{for } i\in\mathcal{N}. 
\end{equation}
In \cite{omic2009protecting}, each individual decides its own recovery rate by seeking treatment, having antidotes to optimize the trade-off between the overhead invested in recovery $c_i \delta_i$ and the penalty of infection $p_{i\infty}$. \cite{omic2009protecting} creates a game with $N$ players whose goals are to minimize $J_i(\delta_i,\delta_{-i}) = c_i \delta_i+p_{i\infty}$. The coupling between individuals' strategies is captured by the limiting behavior of the epidemic model (\ref{eq:steadystateintertwined}). Static game frameworks consider once-and-for-all interventions which cannot be revoked and concern about the long-term outcomes such as the infection risk at the steady-state, or whether the disease will die out eventually. 

Discrete-time stochastic games and differential game frameworks are introduced to capture the transient behavior of the epidemic process and to model adaptive interventions. The difference between Markov game frameworks and differential game frameworks lies in the choice of epidemic models. Discrete-time stochastic game frameworks are built upon stochastic epidemic models such as the stochastic population model (\ref{StochasticPopModel}) and the stochastic network model (\ref{2NExactSISModel}) \cite{eksin2019control,eksin2017disease,lagos2020games}. Differential game frameworks rely on deterministic epidemic models or stochastic epidemic models that use stochastic differential equations to describe the dynamics of the spreading processes \cite{reluga2010game,aurell2020optimal,dashtbali2020optimal,huang2019achieving,huang2020differential,khouzani2011saddle,shen2014differential}. Characterizing a Nash equilibrium over the whole horizon is 
prohibitive for discrete-time stochastic games when the number of individuals increases or the number of stages becomes large. Even structural results are difficult to obtain. Hence, in \cite{eksin2019control,eksin2017disease}, the authors introduce a concept called myopic Markov perfect equilibrium (MMPE). The solution concept MMPE implies the assumption that individuals maximize their current utility given the state of the disease ignoring their future risks of infection and/or future costs of taking interventions in their current decision-making. This is a reasonable assumption considering the computational complexity of accounting for future states of the disease during an epidemic. \cite{lagos2020games} also adopts a similar solution concept where only the current state and the next state of one's infection are considered. For differential game frameworks, the equilibrium can be calculated using the general methods of Isaacs \cite{isaacs1999differential}. Using Pontryagin’s maximum principle, \cite{reluga2010game} studied the differential game of social distancing and the spreading of an epidemic under the equilibrium;  Khouzani et al. \cite{khouzani2011saddle} found the optimal way of dissemination security patches in wireless networks to combat the spread of malware controlled by an adversary; Huang et al. \cite{huang2020differential} characterized the optimal way of reducing connectivity to keep the balance between mitigating the virus and maintaining the economy.  

Repeated game frameworks are used to model seasonal epidemics which appear periodically \cite{breban2007mean,li2017minimizing}. Interventions will be taken repeatedly at each epidemic season. For example, to protect oneself from influenza, one needs to get a flu vaccine each flu season due to the mutation of the virus or the protection time of a vaccine. Individuals adapt his/her behavior based on the cost/payoff incurred last season. Different from differential games and discrete-time stochastic game frameworks, repeated game frameworks do not include the transient behavior of the spreading process \cite{adiga2016delay,huang2019game}.

One common way of individuals making `best-response' decisions that give the best immediate or long-term payoff. This way of decision-making is adopted by differential game, stochastic game, and repeated game frameworks. Another is the use of `imitation' dynamics where individuals copy the behavior that is previously or currently most successful \cite{reluga2006evolving}. The `imitation' dynamics governing the time evolution of the fractions of strategies in the population is similar to the replicator dynamics of evolutionary game theory \cite{poletti2009spontaneous}. Evolutionary game theory has been adopted to model the human behavior of imitating other individuals' successful strategy by many previous studies \cite{hayel2017epidemic,amaral2021epidemiological,zhang2013braess,reluga2006evolving,poletti2009spontaneous}. In the evolutionary game framework for epidemic modeling, the strategy dynamics is coupled with the epidemic dynamics \cite{hayel2017epidemic,zhang2013braess,reluga2006evolving}. The focus of evolutionary game frameworks is on the analysis of the coupled dynamics and  the interpretation of the behavior of these dynamics \cite{amaral2021epidemiological}. A detailed review of this branch of research can be found in \cite{chang2020game}.

Based on the interventions individuals adopt, game-theoretic models in epidemics can be categorized into {vaccination game \cite{adiga2016delay,breban2007mean,li2017minimizing,saha2014equilibria,zhang2013braess,hota2019game}, social distancing game \cite{aurell2020optimal,dashtbali2020optimal,huang2019achieving,huang2020differential,lagos2020games,pejo2020corona,reluga2010game,gosak2021endogenous}, quarantine game \cite{hota2020impacts,amaral2021epidemiological}, and mask wearing game \cite{pejo2020corona} etc}. There are also papers that study the adoption of {preemptive interventions \cite{eksin2017disease,eksin2019control,zhang2013braess} and interventions that changes the recovery rate \cite{hota2018game,xu2015competition}}. Beyond human social networks, many papers studied interventions that can curb the malware or virus spreading in computer networks or wireless communication networks \cite{huang2019game,hayel2017epidemic,huang2019achieving,huang2020differential,khouzani2011saddle,omic2009protecting,shen2014differential,trajanovski2015decentralized,trajanovski2017designing}, such as network re-forming \cite{trajanovski2015decentralized,trajanovski2017designing}, installing security patches \cite{hayel2017epidemic,shen2014differential}, reducing the communication rate \cite{khouzani2011saddle,huang2020differential} and etc. 

If all individuals practice socially distancing, wear masks, stick to stay-at-home orders, the risk of infection and the infection level of the whole population will be reduced significantly. However, there always exist trade-offs and temptations to defect from the regimen. Hand-washing is tedious, wearing a mask is uncomfortable or annoying, socializing is necessary. When it comes to getting a vaccine, people express concerns about safety and side effects. One commonality of most effective interventions usually exhibits the characteristics that if one takes the intervention, he/her needs to pay for all the cost or inconvenience but everyone else in the population will more or less benefit from his/her behavior. This characteristic creates the coupling between individuals. For example, one can enjoy empty streets and markets without having a higher risk of infection if most people stay at home. Those who choose not to get a vaccine effectively reap the benefits of reduced virus transmission contributed by the people who do opt for vaccination. This behavior is referred to as `free-riding' behavior \cite{ibuka2014free}. When a significant number of free rides appear, there will be a collective threat to containing the virus. 

Different interventions induce different costs and provide benefits in different ways. For example, wearing a mask gives immediate protection and is irreversible, hence induces instantaneous cost and benefit. Vaccination creates longer protection yet induces an immediate cost such as making a payment for the vaccine and experiencing side effects. Hence, different models were used to model different interventions. Vaccination games are usually modeled by static games or repeated games due to the irreversible of getting a vaccine \cite{adiga2016delay,breban2007mean,hota2019game,saha2014equilibria}. Revocable interventions such as quarantine, social distancing, wearing a mask are usually modeled by differential games or stochastic games \cite{aurell2020optimal,dashtbali2020optimal,huang2020differential,lagos2020games,reluga2010game}.

Another criterion to classify the literature is to consider the players of the game. Most game-theoretic models in epidemics investigate the interplay between individuals \cite{bauch2004vaccination,bhattacharyya2010game,breban2007mean,chang2020game,dashtbali2020optimal,hayel2017epidemic,hota2020impacts,hota2016interdependent,hota2019game,ibuka2014free,lagos2020games,liu2012impact,reluga2010game,reluga2006evolving,zhang2013braess,eksin2017disease,eksin2019control,chapman2012using,adiga2016delay,huang2020differential,li2017minimizing,saha2014equilibria,trajanovski2015decentralized,trajanovski2017designing}. Individuals can be completely selfish who maximize their own payoff via optimizing their own payoff or imitating the most successful individuals \cite{adiga2016delay,bauch2004vaccination,bhattacharyya2010game,breban2007mean,chang2020game,dashtbali2020optimal,hayel2017epidemic,hota2020impacts,hota2016interdependent,hota2019game,ibuka2014free,lagos2020games,liu2012impact,reluga2010game,reluga2006evolving,zhang2013braess}. Some papers incorporate the effect of altruism into their game-theoretic models in which individuals are not completely selfish and care about the well-being of their neighbors \cite{eksin2017disease,eksin2019control,chapman2012using}. It is shown by Eksin et al. \cite{eksin2017disease} that a little empathy can significantly decrease the infection level of the whole population. The results in \cite{chapman2012using} by Chapman et al. show that the central planner should promote vaccination as an act of altruism, thereby boosting vaccine uptake beyond the Nash equilibrium and serving the common good. Several papers examine the inefficiency of selfish acts of individuals and the inefficiency is quantified by the price of anarchy \cite{adiga2016delay,huang2020differential,li2017minimizing,saha2014equilibria,trajanovski2015decentralized,trajanovski2017designing}. Results from these studies demonstrate that individuals' selfishness becomes a big hurdle to fight against infectious diseases. Hence, some papers introduce the role of central authorities and study how central authorities should create incentives/penalties to achieve social optimum \cite{aurell2020optimal,farhadi2019efficient,pejo2020corona}. Another strain of research focuses on the interplay between a central authority and an adversary \cite{khouzani2011saddle,shen2014differential,xu2015competition,zhang2019differential}. The adversary aims to maximize the overall damage inflicted by the malware and the central authority tries to find the best counter-measure policy to oppose the spread of the infection. The conflicting goals between the players are usually captured by a zero-sum dynamic game.

\subsection{A Fine-Grained Dynamic Game Framework for Human-in-the-Loop Epidemic Modelling}\label{subsec:FineFrainedFramwork}

In the existing literature, there is no consensus on which game-theoretic framework to study human-in-the-loop epidemics. The integration between game-theoretic models and epidemic models is done on a case-by-case basis depending on the players involved, what interventions people take(see Section~\ref{subsec:interventions}), when interventions are taken (see Section~\ref{sebsec:whenInterventions}), what epidemic models modelers choose (see Section~\ref{subsec:EpidemicModels}), and the underlying network structure. Here, we present a fine-grained dynamic game framework to describe the essence of human-in-the-loop epidemic modeling.

We consider the following discrete-time Markov game with $N$ players.

\begin{itemize}
    \item[\textbullet] \textit{Players}: We consider a population of $N$ individuals denoted by $\mathcal{N}$ and a central authority.
    \item[\textbullet] \textit{The individual state space}: Each individual has a state from the finite set $\mathcal{S}$. The state indicates the health status of an individual. Elements in $\mathcal{S}$ may include susceptible state, infected state, recovered state, and/or quarantined state etc., depending on the compartment model being used and how interventions are modeled. The state of individual $i$ at time $t$ is denoted by $X_i(t) = \{1,2,\cdots, \vert \mathcal{S} \vert \}$. For example, if we consider an $SIS$ model where individuals decide whether to get vaccinated, the state space $\mathcal{S}$ contains three elements $\{1,2,3\}$, with $X_i(t) = 1$ ($X_i(t)=2,X_i(t)=3$) meaning individual $i$ is susceptible (infected, vaccinated respectively).
    
    One can also introduce the concept of type in game theory into the human-in-the-loop framework to capture different social, political, and other demographic groups.
    
    \item[\textbullet] \textit{The population profile}: The global detailed description of the population's health status at time $t$ is $\bar{X}(t) = (X_1(t),\dots,X_N(t))^T\in \mathcal{S}^N$. The population profile is denoted by $M(t) = (M_1(t),M_2(t),\dots,M_{|\mathcal{S}|}(t))^T$, where $M_s(t) = (1/N) \sum_{i=1}^N \mathbbm{1}_{X_i(t)=s}$, indicating the proportion of individuals in state $s$. A central authority cares about the well-being of the whole population, hence pays attention only to the population profile $M_s(t), s\in\mathcal{S}$.
    \item[\textbullet] \textit{The individual action space}: Let $\mathcal{A}$ denote a set of possible actions individuals can take to combat the virus when his/her state is $s$. Depending on which intervention is studied (see Section~\ref{subsec:HowInterventionsModelled}), $\mathcal{A}$ can be either finite or continuum. At every time $t$, an individual chooses an action $A_i(t) \in\mathcal{A}$. The action profile of the whole population is denoted by $\bar{A}(t)=(A_1(t),\dots,A_N(t))^T$. The action set can be state-dependent if necessary.
    \item[\textbullet] \textit{The transition kernel}: The epidemic process $(\bar{X}(t))_{t\in\mathbb{N}}$ is a Markov Process once the sequence of actions taken by individuals is fixed. Let $\Gamma$ be the transition kernel, namely $\Gamma$ is a mapping $\mathcal{S}^N\times \mathcal{S}^N \times \mathcal{A}^N \rightarrow [0,1]$. Given the population infection profile $\bar{X}(t)$ at time $t$, if individuals take their actions $(A_i(t))_{i\in\mathcal{N}}$, then at time $t+1$, the global detailed description of the population's health status follows the distribution:
    $$
    \begin{aligned}
    \mathbb{P}(\bar{X}(t+1)=(s_1,\dots,s_N)^T\vert \bar{X}(t) &= (x_1,\dots,x_N)^T,\bar{A}(t)=(a_1,\dots,a_N))\\
    &=\Gamma(x_1,\dots,x_N,s_1,\dots,s_N,a_1,\dots,a_N).
    \end{aligned}
    $$
    The transition kernel is decided by the epidemic models (see Section~\ref{subsubsec:StochasticModels}) and the interventions the actions represent (see Section~\ref{subsec:HowInterventionsModelled}). For example, the transition kernel can be constructed using the SIS networked stochastic model (\ref{2NExactSISModel}) or other epidemic models. For example, if the actions of individual $i$ include $a_i\in\{0,1\}$ with $a_i=0$ representing individual $i$ is self-quarantined and $a_i=1$ representing individual $i$ staying normal, then the transition kernel can be constructed by
    \begin{equation}\label{eq:exampTransitionKernel}
    \begin{aligned}
        &\mathbb{P}(X_i(t+1) =0\vert\ X_i(t)=1, \bar{X}(t), A_i(t) = a_i) = \delta \Delta t + o(\Delta t),\\
        &\mathbb{P}(X_i(t+1) = 1\vert\ X_i(t)=0,\bar{X}(t), A_i(t) = a_i) = \beta a_i \sum_{j=1}^N a_{ij} X_j(t) \Delta t +  o(\Delta t),
    \end{aligned}
    \end{equation}
    for $i\in \mathcal{N}$. As we can see if all individuals brace for the epidemic by quarantining, i.e., $A_i(t)=0$ for all $i\in\mathcal{N}$ for some $t\in[t_1,t_2]$, the spreading will slow down at a fast rate decided by $\delta$. Not every individual has the inventive to do so.
    \item[\textbullet] \textit{Costs}: The costs come from two sources: one is from the risk of catching the virus, another is from the interventions being taken. Most interventions/measures come with either monetary costs or inconvenience. Hand-washing is tedious, wearing a mask is uncomfortable or annoying, socializing is necessary. The instant cost at time $t$ for individual $i$ hence depends on his/her health status (state), the interventions he/she takes (action), and/or the states of other individuals. Formally, 
    $$
    G_i(t) = g^s_i(X_i(t),A_i(t)) + g_i^f(X_{\mathcal{N}_i},A_{\mathcal{N}_i}) + g_i^a(\bar{X}(t),\bar{A}(t)),
    $$
    in which there exits three levels of altruism. The cost function $g_i^s$ captures the selfishness of individual $i$, where the superscript $s$ means selfishness. If the second cost function $g_i^f$ is added, it means individual $i$ care about his/her neighbors (fiends, family members). The superscript $f$ of $g_i^f$ means friends or family.  The highest level of altruism is captured by $g_i^a$ meaning individual $i$ cares about every individual in the population. For a completely selfish individual $g_i^f(\cdot,\cdot) \equiv 0$ and $g_i^a(\cdot,\cdot) \equiv 0$. Modellers can also make $g_i^s$, $g_i^f$, and $g_i^a$ time dependent if necessary.
    \item[\textbullet] \textit{Information and Strategies}: The information set of individual $i$ at time $t$ is denoted by $\mathcal{I}_i(t)$. Individual may not every detail about the whole population. He/she may only know his own health status and the population profile broadcasted by the central authority, i.e., $\mathcal{I}_i(t) = \{X_i(0),\dots,$ $X_i(t),M(0),\dots,M(t)\}$. Or individual $i$ may only know information about his neighboring individuals, which gives $\mathcal{I}_i(t)=\{X_{\mathcal{N}_i}(0),\dots,$ $X_{\mathcal{N}_i}(t)\}$. Individuals make decisions based off of the information available to them. The rules individuals follow to make decisions is called strategies. The strategy of individual $i$, $\sigma_i$, is a map from the information space and the time space to his/her action space, meaning that the action of individual $i$ is chosen as $A_i(t) = \sigma_i(\mathcal{I}_i(t),t)$. 
\end{itemize}

\begin{remark}
When there is a presence of a central authority, the central authority cares about the population profile instead of the health status of a particular individual. For example, the goals of the central authority might be to suppress the proportion of infected individuals,  reduce the death toll on the general public, and boost up the uptake of vaccines. These metrics can all be reflected in the population profile $M(t)$ for $t=0,1,2,\cdots$. Hence, the cost function of the central authority can be a function of $M(t)$, i.e., $G_c(t) = g_c(M(t))$. There are two paths the central authority can follow to achieve its goals. The first is designing a penalty/reward function $g_i^p(X_i(t), A_i(t))$ that rewards (punish) individuals who comply with (violate) the suggested rules such as social distancing, quarantine, or getting vaccinated. The second is through information design such as promoting altruism, raising awareness, health education, etc.
\end{remark}

The game between individuals unfolds over a finite or infinite sequence of stages, where the number of stages is called the horizon of the game. Infinite horizon game models epidemics that persist for decades \cite{reluga2010game}. Some papers consider finite horizon game where the terminal time is decided by when the vaccines are widely available \cite{reluga2006evolving,huang2020differential,chang2020game}. The overall objective, for each individual, is to minimize the expected sum of costs he/she receives during the epidemic. 

Solving such a fine-grained stochastic game is difficult, if not prohibitive, under general solution concepts. The difficulties emerge from three facts. The first is that as the number of individuals $N$ increases, the size of the state space $|\mathcal{S}|^N$ for the global state $\bar{X}$ increases exponentially. In the human population, the number of individuals in a community ranges from thousands to millions. Hence, analyzing such an enormous number of individuals under the fine-grained dynamic game framework becomes impossible. The second is that individuals do not know the exact information of the whole population $\bar{X}(t)$ for every $t$. For some epidemics, it is difficult for individuals to know their own state due to the fact that some infectious diseases do not cause symptoms for some individuals or cause common symptoms that are shared with other diseases. 
Also to gather information regarding the population profile $M(t)$ for each time $t$ and broadcast it to the individuals, the central authority needs to arrange large-scale surveys, polls, and diagnostic tests on a daily or weekly basis. Even so, the population profile $M(t)$ can only be estimated using gathered data. This partial information situation creates a partially observable stochastic game. It is intractable to compute Nash or other reasonable strategies for such partially observable stochastic games in most general cases \cite{horak2019solving}. The third is if the transition kernel is described by a networked stochastic epidemic model such as (\ref{eq:exampTransitionKernel}), the state dynamics of one individual is coupled directly or indirectly with every other individual through the underlying network. The fact makes it impossible to obtain an appropriate strategy by decoupling \cite{eksin2017disease}. Hence, we propose this fine-grained stochastic game framework to describe the quintessences of the integration model into the epidemic spreading process, with no intention to solve it.

Existing literature usually proposes less fine-grained game-theoretic frameworks in order to obtain meaningful results that help understand the spreading of epidemics under human responses. These papers generalize or simplify the fine-grained dynamic game mainly in three ways. One is to use mean-field techniques by assuming the transition probability of each individual's state only couples with the population profile $M(t)$ and some 
indistinguishability assumptions \cite{tembine2009mean,gast2012mean,lee2021controlling,tembine2020covid, reluga2010game}. The second is to study a simplified solution concept where each individual considers costs over only a limited number of stages \cite{eksin2019control,eksin2017disease,lagos2020games}. The third is to consider a continuous $N$-intertwined epidemic model (\ref{MieghemSIS}), rather than a stochastic one, where a dynamic game is built upon \cite{huang2020differential} or to apply mean-field techniques in a homogeneous epidemic model such as the Kermack and McKendrick model (\ref{KMSISModel}) \cite{reluga2010game}. In the next subsection, we present three representative papers that proposed simplified frameworks to deliver meaningful results by leveraging the above-mentioned methods. These frameworks choose totally different epidemic models and have their unique ways of integrating game-theoretic decision-making into these epidemic models.

\subsection{Social Distancing Game with Homogeneous SIR Epidemic Model \cite{reluga2010game}}\label{subsec:reluga2010game}

In \cite{reluga2010game}, T. Reluga studies the effect of social distancing on the spreading of SIR type of infectious diseases.  The author uses a homogeneous deterministic epidemic model: the Kermack and McKendrick model (\ref{KMSISModel}). A differential game framework is proposed in which the interplay between individuals is simplified as the interaction between a specific individual and the aggregate behavior of other individuals.  

\subsubsection{Modelling} \label{subsubsec:modellingReluga}

\textit{How social distancing is modeled}: Let $a$ be one specific individual's strategy of daily investment in social distancing. The population strategy $a_s$ is the aggregate daily investment in social distancing by the population. Borrowing the idea from mean-field games, in the limit of infinitely large populations, i.e., $N\rightarrow \infty$, $a$, and $a_s$ are independent strategies because changes in one individual's behavior will have a negligible effect on the average behavior. The effectiveness of investment in social distancing is captured by $\sigma(a_s)$, which is the infection rate given an aggregate investment in social distancing practices. Without loss of generality, we set $\beta(0)=1$ when there is no investment. To model the diminishing returns with increasing investment, the author assumes that $\sigma(a_s)$ is convex and given by
$$
\sigma(a) = \frac{1}{1+ma},
$$
with the maximum efficiency of social distancing $\sigma'(0)=-m$. 

\textit{Epidemic Models}: Epidemic usually start with one or a few infected cases, so $I(0)\approx 0$. The macroscopic behavior of the spreading process under the aggregate social distancing  investment $a_s$ can be captured by a normalized SIR version of the Kermack and McKendrick epidemic model:
\begin{equation}\label{eq:KMinReluga}
\begin{aligned}
\dot{S} &= -\sigma(a_s)SI\\
\dot{I} &=  \sigma(a_s)SI - I(t),\\
\dot{R} &= I(t),
\end{aligned}
\end{equation}
where the infection rate and the recovery rate are normalized in order to focus on the effect of social distancing practices on the spreading process. 

\textit{The cost function:} The total cost of the epidemic to the population, $J$, includes the daily costs from infection $a_s S$, the daily costs of infection $I$:
\begin{equation}\label{eq:communitycostreluga}
J = -\int_{0}^{t_f} (a_s S + I)e^{-ht}dt - \frac{I(t_f) e^{-h t_f}}{1+h} 
\end{equation}
where $e^{-ht}$ is a discount term with $h$ being the discount rate and the last term is a salvage term representing the cumulative costs associated with individuals are sick at the time mass vaccination occurs ($t_f$).

\textit{The evolution of individual states:} The premise of the game is that at each point in the epidemic, individuals can choose to pay a cost associated with social distancing in exchange for a reduction in their risk of infection. Let $\mathbf{p(t)}\in \Delta^3$ be the probabilities 
that an individual is in the susceptible, infected, or recovered state at time t. The probabilities $\mathbf{p}(t)$ evolve according to the Markov process
\begin{equation}\label{eq:individualRiskinreluga}
\dot{\mathbf{p}} = P(t;a)\mathbf{p},
\end{equation}
where $a$ is the individual's daily investment in social distancing and the transition-rate matrix
$$
P(t;a) = \begin{bmatrix}
-\sigma(a) I & 0 & 0\\
\sigma(a) I & -1 & 0\\
0 & 1 & 0
\end{bmatrix}.
$$
The coupling between the population profile $(S,I,R)$ and the individual profile $\mathbf{p(t)}$ is described by the two processes (\ref{eq:KMinReluga}) and (\ref{eq:individualRiskinreluga}). The individual risk of infection depends on his/her investment in social distancing $a$ and the infection level of the whole population $I$.

\textit{The values of states:} Using the ideas of Isaacs \cite{isaacs1999differential}, we calculate  expected present values of each state at each time, conditional on the
investment in social distancing. The expected present value is
average value one expects after accounting for the probabilities of
all future events, and discounting future costs relative to immediate
costs. Let $\mathbf{V}(t;a,a_s) = \left(V_S(t;a,a_s),V_I(t;a,a_s) ,V_R(t;a,a_s) \right)$ denote the expected present values with $V_S(t; a,a_s)$, $V_I(t; a,a_s)$, and $V_R(t; a,a_s)$ representing the expected present values of being in the susceptible, infected, or removed state at time $t$ when using strategy $a$ in a population using strategy $a_s$. The expected present values $\mathbf{V}$ eolves according to the adjoint equations
$$
-\dot{\mathbf{V}} = (P(t;a)^T - h I_d) \mathbf{V} + \mathbf{v},
$$ 
where $\mathbf{v}^T(t;a) =(-\sigma(a),-1,0)$ incorporates
the individualized cost of (\ref{eq:communitycostreluga}) into the expected present values $\mathbf{V}(t;a,a_s)$. Since the dynamics of $\mathbf{V}$ is independent of $R$, there is no need to consider recovered individuals further. Further simplifying the dynamics of $\mathbf{V}$ by taking $h=0$ (not discount) and $V_I=-1$ (fixed expected value at infected state), one obtains
\begin{equation}\label{eq:ValueEvolutionReluga}
    -\dot{V}_S = -(1 + V_S)\sigma(a) I -a,
\end{equation}
which evolves backward in time with boundary condition $V_S(t_f) =0$. If everyone else invest heavily in social distancing, the individual can become a free rider that earns the benefit (value) without having to invest too much in social distancing because the infection level $I$ will remain low. To find a balanced social distancing strategy (Nash strategy), one can simply focuses on (\ref{eq:KMinReluga}) and (\ref{eq:ValueEvolutionReluga}).

\subsubsection{Analysis}

To find a balanced strategy is to find the best strategy to play, given that all the other individuals are also attempting to do so. Given such context, a Nash equilibrium solution becomes an appropriate solution concept for the differential game formulated in Section~\ref{subsubsec:modellingReluga}.
\begin{definition}
Given the expected value $V_S(t;a,a_s)$ in the susceptible state and the associated population level spreading dynamics (\ref{eq:KMinReluga}), $a^*$ is a Nash equilibrium if for any possible strategy $a$, $V_S(0;a,a^*) \leq V_S(0;a^*,a^*)$.
\end{definition}
A Nash equilibrium is a subgame perfect equilibrium if it is also a Nash equilibrium at every state the system may pass through. Indeed, the Nash equilibrium can be obtained by finding the investment that maximizes the rate of increase in the individual’s expected value. 
\begin{lemma}\label{lemma:NecConReluga}
If $a^*(V_S,I)$ is a subgame perfect equilibrium, then it satisfies the maximum principle
$$
a^*(V_S,I) = \arg\max_{a\geq 0} -(1+V_S)\sigma(a)I - a,
$$
when $a_s = a^*$ everywhere.
\end{lemma}
One can solve for $a^*(V_s,I)$, if $\sigma(a)$ behaves well.
\begin{theorem}\label{theorem:inregula}
If $\sigma(a)$ is differentiable, decreasing, and strictly convex, then $a^*$ is uniqely defined by the relations
\begin{equation}\label{eq:socialdistancingrule}
\begin{cases}
a^* =0\ \ \ \ \ \ \textrm{if } -\sigma'(0)I(1+V_S)\leq 1,\\
-\sigma'(a^*) I (1+V_S)=1,\ \ \ \textrm{otherwise}.
\end{cases}
\end{equation}
\end{theorem}
From Theorem~\ref{theorem:inregula}, one knows that at the Nash equilibrium, whether an individual invests in social distancing depends on the maximum efficiency of social distancing $\beta(0)'=m$,  the current infection level of the population $I$, and the expected value in susceptible state $V_S$. {When an individual does invest in social distancing, an individual tends to invest more if any of the following values is higher:  the efficiency of social distancing, the current infection level $I$, the value of the susceptible state $V_S$.} 

\subsubsection{Highlighted Results}

\begin{figure}
  \includegraphics[width=0.7\textwidth]{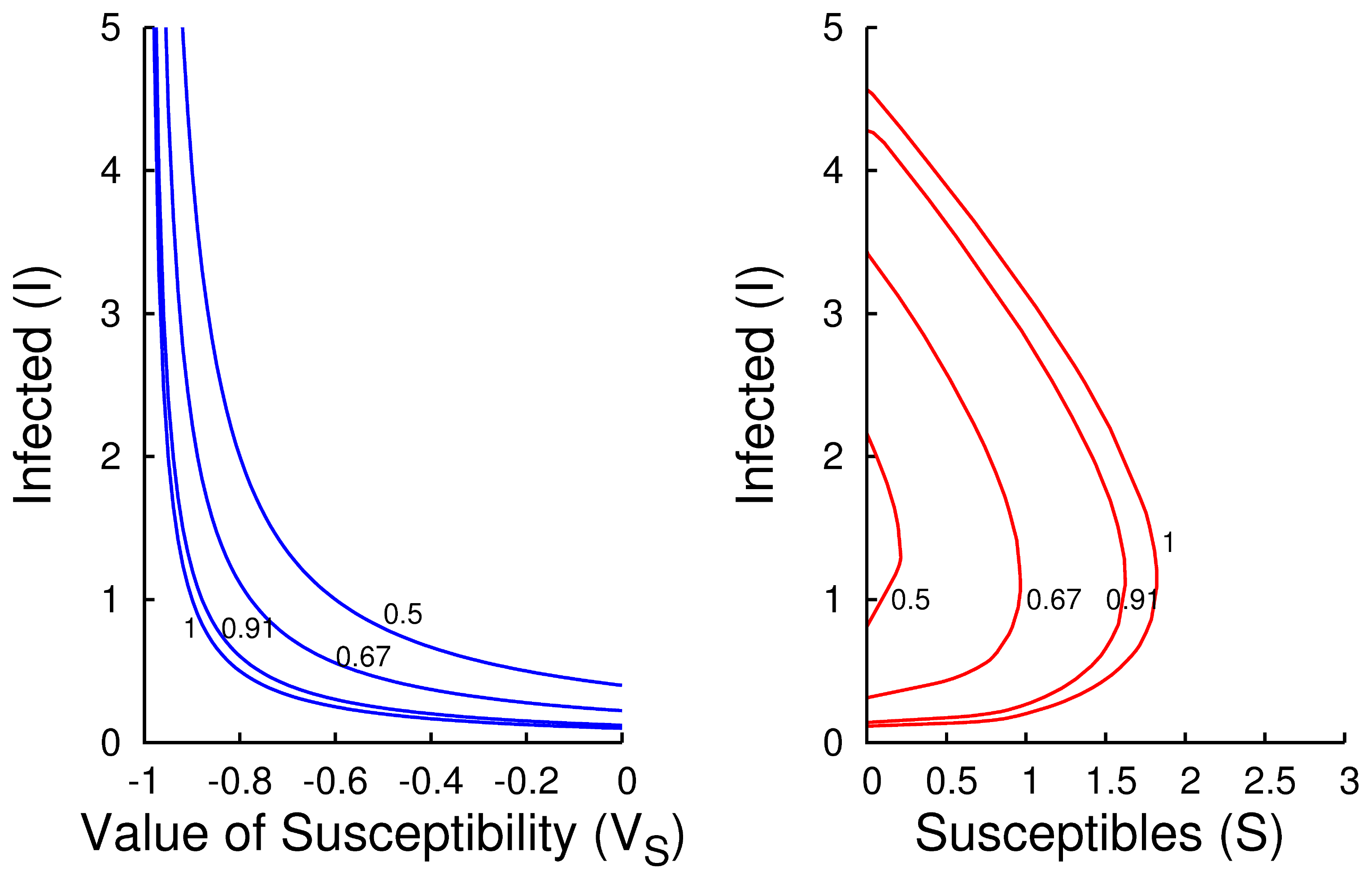}
\caption{\cite{reluga2010game} Contour plots of relative risk surface for equilibrium
strategies with parameter $m=10$.}
\label{fig:ContourPlotsRelativeRisk}      
\end{figure}

{The authors investigate the instantaneous behavior $a^*$ given the expected value in susceptible state $V_S$, the infection level $I$, and the susceptible level $S$. Such results can be computed by solving (\ref{eq:socialdistancingrule}) and the results are shown in Figure~\ref{fig:ContourPlotsRelativeRisk}. From (\ref{eq:socialdistancingrule}), we know that there are two types of equilibrium strategies including no investment in social distancing ($a^* = 0, \sigma(a^*)=1$) and positive investment in social distancing ($a^*>0,\sigma(a^*)<1$). The first figure in Figure~\ref{fig:ContourPlotsRelativeRisk} is a contour plot in $V_S \times I$ surface about the relative risk $\sigma(a^*(V_S,I))$. The values attached to the blue lines represent what the relative risks $\sigma(a^*(V_S,I))$ are at the corresponding coordinates $(V_S,I)$. The line with value $1$ attached to it separates the region where the equilibrium strategy will include no investment from the region where the equilibrium strategy requires investment in social distancing. The feedback form of equilibrium strategies, transformed from $(V_S,I)$ coordinates to the $(S,I)$ coordinates of the phase-space is represented with another contour plot in the second figure of Figure~\ref{fig:ContourPlotsRelativeRisk}.}

\textit{The last few `survivors' tend to social distance more}: As we expect, the left figure shows that a larger value of the susceptible state $V_S$ induces a greater instantaneous social distancing.  From the right figure, one can see that as the number of susceptible individuals increases, the investment in social distancing decreases, hence the individual's infection rate increases with less social distancing. One can also see that when only a small portion of the population remains susceptible, the biggest investments in social distancing happens. That means the last few 'survivors' tend to social distance to brace for the infection.

\begin{figure}
  \includegraphics[width=0.9\textwidth]{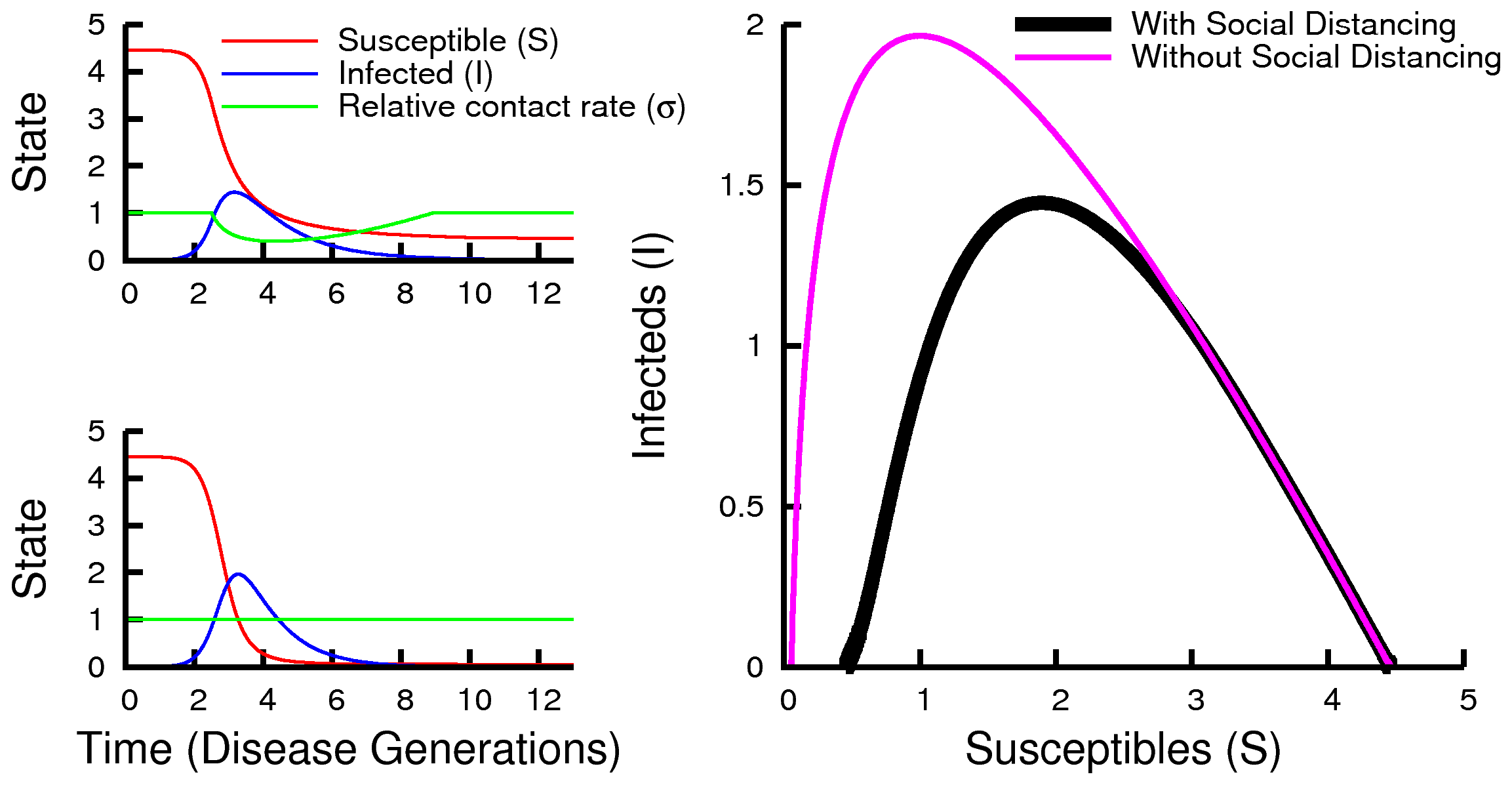}
\caption{\cite{reluga2010game} The spreading dynamics with equilibrium social distancing and without social distancing with parameters $R(0)=4.46,m=10$.}
\label{fig:DynamicsWithAndWithoutSD}      
\end{figure}

Two scenarios are investigated in \cite{reluga2010game}. The first is the infinite horizon differential game that gives the equilibrium behavior when there is never a vaccine and the epidemic spreads until its natural end. The second is the finite-horizon problem that studies the individual behavior in equilibrium when there will be a vaccine introduced at time $t_f$. For the infinite-horizon case, the epidemic spreading dynamics under the social distancing equilibrium and in the absence of social distancing is plotted in Figure~\ref{fig:DynamicsWithAndWithoutSD}.

\textit{Social distancing occurs later but ends sooner than the wide spreading of epidemics:} As we can see from the top left figure in Figure~\ref{fig:DynamicsWithAndWithoutSD}, under equilibrium social distancing, social distancing is never used until part-way into the epidemic and ceases before the epidemic fully dies out. That means at the beginning of the epidemic, individuals will not be alert to take any interventions until the epidemic prevails. And social distancing practices are going to be lifted before the epidemic completely ends when the situation gets better.

\textit{Social distancing
leads to a smaller epidemic but prolongs the epidemic:} When comparing the time series data on the top left figure and its counterpart on the bottom left figure of Figure~\ref{fig:DynamicsWithAndWithoutSD}, one can observe that social distancing reduces the scale of the epidemic and prolongs the prevalence of the epidemic. Even though social distancing prolongs the epidemic, practicing social distancing is still important since it helps `flatten the curve'.  {``flattening the curve'' reduces the number of cases that need intensive care at any given time, giving health-care workers, hospitals, police, schools, and vaccine-developers time to prepare and respond without becoming overwhelmed. Essential facilities (e.g., hospitals and schools) can still function normally with a tolerable level of infection. Slowing and spreading out the tidal wave of cases will save lives. Flattening the curve keeps society going.}

\begin{figure}
  \includegraphics[width=0.9\textwidth]{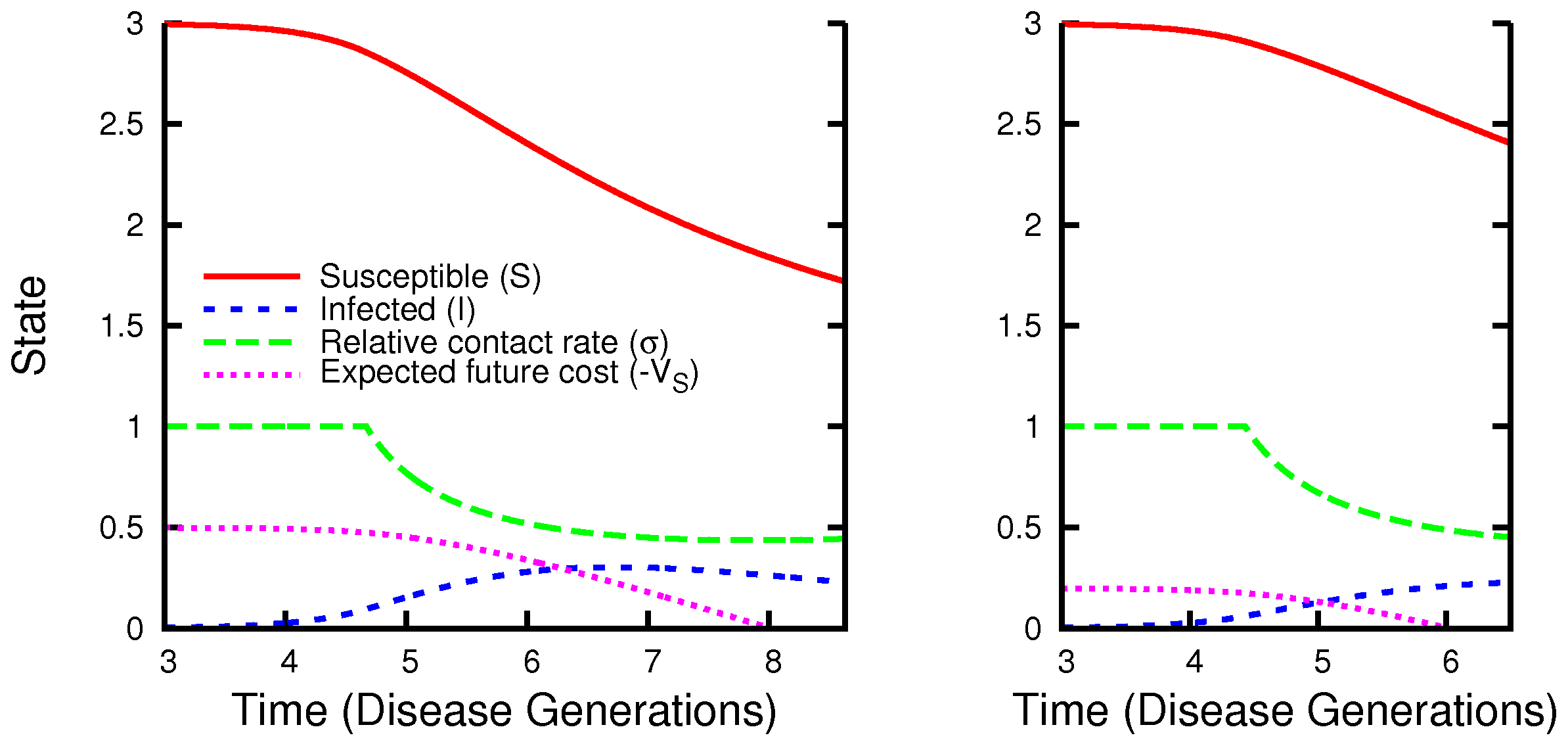}
\caption{\cite{reluga2010game} The spreading dynamics when universal vaccination occur after a fixed time with parameters $R(0)=3,m=20,I(0)= 3\times 10^{-6}$.} 
\label{fig:FiniteTimeSpreadingDynamics}     
\end{figure}

Now let's shift the focus to the finite-horizon problem where vaccines are universally available at a given time $t_f$. As is shown in Figure~\ref{fig:FiniteTimeSpreadingDynamics}, at an equilibrium, social distancing will last until the very time when vaccines are universally available. When the wide availability of vaccines arrives sooner, social distancing begins sooner. When the vaccine becomes available at $t_f = 8.6$ (see the left plot in Figure~\ref{fig:FiniteTimeSpreadingDynamics}),  individuals save $50\%$ of the cost of infection per capita by practicing social distancing. When the vaccine becomes available earlier, say when $t_f = 6.5$, $80\%$ of the cost of infection can be saved per capita.

\textit{Social distancing enlarge the window of opportunity during which mass vaccine can reduce the cost of the epidemic:} The earlier a vaccine becomes available, the less the whole society suffers. If a vaccine becomes available at the late stage of the epidemic when most individuals are recovered, the vaccine won't help much reduce the transmission. There exists a limited window during which large-scale vaccination can effectively cut down the cost of infection at the population level. Numerical results in \cite{reluga2010game} show that equilibrium social distancing can extend this limited window of opportunity.

\subsubsection{Discussions}

The modeling in \cite{reluga2010game} unravels the complexity of the fine-grained dynamic framework from several aspects. The first is the use of a homogeneous SIR deterministic epidemic model: the Kermack–McKendrick SIR
model, in which the epidemic process is described by two ordinary differential equations ($R$ can be expressed as $R=1-S-I$). The effect of social distancing on the spreading process is captured by a scalar function $\sigma(a_s)$ which is homogeneous to all individuals. The spreading process enjoys a decreased  infection rate as the population invests more in social distancing.
The use of such a homogeneous epidemic model is a double-edged sword. On one hand, homogeneous epidemic models make analytical results more attainable, but on the other side, homogeneous epidemic models need the assumption that the population is homogeneous and strongly mixed. However, we know that the contact patterns among individuals are highly structured, with regular temporal, spatial, and social correlations.  The second is the decoupling of the direct connection between an individual's strategy $a$ and the aggregate strategy of the population $a_s$. This allows one's risk of infection to depend only on the infection level of the population $I$, which implicitly depends on $a_s$.

Realistically, mass vaccination cannot happen overnight as is assumed in the paper. Vaccination is usually rolled out continuously as it is proved to put into use. This effect can be incorporated into the model by considering a time-dependent forcing. In this game, the individual has complete information about the epidemic including the expected value and the infection level of the whole population. However, in reality, incomplete information (biased or inaccurate information) may drive human behavior away from the equilibria obtained in this paper.

\subsection{The Power of Empathy: A Markov Game under the Myopic Equilibrium \cite{eksin2017disease}}

In \cite{eksin2017disease}, Eksin et al. have proposed a Markov game framework using the contact network stochastic epidemic model \cite[Ch. 17]{newman2010networks} in which healthy individuals utilize protective measures to avoid contracting a disease and sick individuals utilize preemptive measures out of empathy to avoid spreading a disease. A solution concept, called the myopic Markov perfect equilibrium (MMPE), is introduced to model human behaviors, which also makes theoretical results attainable for such a framework. Eksin et al. have shown that there is a critical level of empathy by the sick individuals above which the infectious disease die out rapidly. Further, they show that empathy among sick individuals is more effective than risk-aversion from healthy individuals.

\subsubsection{Modelling}

\textit{The epidemic model}: Eksin et al. considers a networked SIS  stochastic epidemic model, which is a variant of the networked stochastic model (\ref{2NExactSISModel}). An individual $i$ in the population $\mathcal{N}$ susceptible ($X_i(t) = 0$) or infected and infectious ($X_i(t)=1$) at any given time $t=1,2,\cdots$. A global detailed description of the population is denoted by $\bar{X}(t) = (X_1(t),\cdots, X_N(t))^T$.

The transition kernel $\Gamma$ can be specified by $p_{01}^i(t)$ and $p_{10}^i(t)$ for $i\in\mathcal{N}$ and $t=1,2,3,\cdots$. $p_{01}^i(t)$ is  the probability that individual $i$ is susceptible at time $t$, but gets infected at time $t+1$ and 
\begin{equation}\label{eq:transProb1Eksin}
p_{01}^i(t) \coloneqq \mathbb{P}(X_i(t+1)=1\vert X_i(t)=0) = 1 - \prod_{j\in\mathcal{N}_i}(1-\beta a_i(t)a_j(t) X_j(t)),
\end{equation}
where $\beta\in(0,1)$ is the infection rate, $0\leq a_i \leq 0$ is the action of individual $i$, and the action of its neighboring individual $j$ is $0\leq a_j \leq 1$ for $j\in\mathcal{N}_i$.

\textit{How protective and preemptive measures are modeled:} Each term $1-\beta a_i(t)a_j(t) X_j(t)$ is the probability that the individual $i$ is not infected by neighbor $j$. If either individual $i$ or his/her neighbor $j$ takes protective and preemptive measures such as wearing masks, practice social distancing, or other measures, the probability that individual $i$ is infected by neighbor $j$ will decrease. An extreme case is either $a_i =0$ or $a_j=0$ under which individual $i$ will never be infected by his/her neighbor $j$. The product
of all the terms $\prod_{j\in\mathcal{N}_i}(1-\beta a_i(t)a_j(t) X_j(t))$ is the probability that the individual is not infected by any  interactions. $p_{10}^i(t)$ is the probability that individual $i$ is infected at time $t$ but recovered at time $t+1$. This probability is equal to the inherent recovery rate of the disease $\delta\in(0,1)$, i.e.,
\begin{equation}\label{eq:transProb2Eksin}
p_{10}^i(t) \coloneqq \mathbb{P}(X_i(t+1) =0 \vert X_i(t)=1) = \delta.
\end{equation}

\textit{The payoff function:} Individuals make their decisions based on the trade-off between the risk of contracting the virus and the costs of taking protective and preemptive measures. Different from other studies in which individuals are completely selfish \cite{adiga2016delay,bauch2004vaccination,bhattacharyya2010game,breban2007mean,chang2020game,dashtbali2020optimal,hayel2017epidemic,hota2020impacts,hota2016interdependent,hota2019game,ibuka2014free,lagos2020games,liu2012impact,reluga2010game,reluga2006evolving,zhang2013braess}, Eksin et al. considers a sense of altruism among individuals. Individuals are concerned not only about getting infected themselves but also about infecting others in their neighborhood. An individual's payoff at time $t$ is a weighted linear combination of these considerations:
{\begin{equation}\label{eq:payofffunEksin}
\begin{aligned}
&g_i(a_i,a_{\mathcal{N}_i}, {X}_{\mathcal{N}_i}(t))\\
=& a_i \beta \left[ c_0 - c_1(1-X_i(t)) \sum_{j\in\mathcal{N}_i}a_j X_j(t) + c_2 X_i(t) \sum_{j\in\mathcal{N}_i} a_j(1-X_j(t)) \right],
\end{aligned}
\end{equation}}
where $c_0,c_1.c_2$ are fixed weights. The first term inside the square bracket is the payoffs of not taking any measures including socialization benefits, convenience benefits, and economic benefits etc. The second term captures the risk aversion of susceptible individuals. The risk comes from contacting with infectious neighbors who fail to take serious measures to protect others. The third term is the empathy term that quantifies the risk of infecting others. Hence, the weights $c_0,c_1$, and $c_2$ are referred to as socialization, risk aversion, and empathy constants.

The payoff function $g_i$ is a bilinear function of $a_i$ and $a_j$ for $j\in\mathcal{N}_j$. That means given the actions of individual $i$'s neighbors $a_{\mathcal{N}_j}$ and their health status $X_{\mathcal{N}_j}(t)$, to maximize his/her payoff, he/she needs to decide whether to resumes normal activity ($a_i=1$) or self-isolates ($a_i=0$) depending on the sign of expression inside the square bracket. If the expression is positive, individual $i$ prefers to resume normal. Otherwise, self-isolation is the best choice.

The payoffs of the neighbors of individual $i$ depend on the actions of their own neighbors. This means if the underlying network $\mathcal{G}$ is connected, the payoff profile of the population couples the actions of all individuals. Hence, individuals need to reason about the interaction levels of their neighbors in their decision-making. Such individual reasoning can be modeled using game theory.

\subsubsection{Analysis}

Obtaining an analytical solution such as the Nash equilibrium is difficult if one considers accumulative payoffs over a finite period of time or an infinite horizon. Here, Eksin et al. have considered a solution concept called the myopic Markov perfect equilibrium (MMPE). 
\begin{definition}
The strategy of individual $i$ at time $t$, denoted by $\sigma_i$, is a mapping from the state $\bar{X}(t)$ to the action space $[0,1]$, i.e., $a^*_i(t) = \sigma_i(\bar{X}(t))$. A strategy profile $\sigma\coloneqq\{\sigma_{i}\}_{i\in\mathcal{N}}$ is called an MMPE strategy profile if 
$$
g_i(a^*_i,a^*_{\mathcal{N}_i}, {X}_{\mathcal{N}_i}(t)) \geq g_i(a_i,a^*_{\mathcal{N}_i}, {X}_{\mathcal{N}_i}(t)),
$$
holds for any $a_i\in[0,1]$ and for all $t=1,2,\cdots$, $i\in\mathcal{N}$ under the Markov process described by (\ref{eq:transProb1Eksin}) and (\ref{eq:transProb2Eksin}).
\end{definition}
The use of MMPE profile carries two implied assumptions. One is the assumption that individuals' actions depend only on the payoff relevant state of the disease. Whether the class of Markovian strategies contains the Nash strategy for all possible strategies is not discussed in \cite{eksin2017disease}. Another is the assumption that individuals make decisions considering the current instantaneous payoff only. Under the assumption of myopic strategies,  individuals do not foresee their future risks of infection or infecting others in their decision-making. 

The computation of the MMPE strategy profile  involves only one stage of the payoff. So, it is more tractable than computing the Nash strategies that consider the accumulative payoffs with states evolving from time to time. Indeed, Eksin et al. show that there exists at least one such strategy profile for the bilinear game captured by (\ref{eq:payofffunEksin}). The proof of existence is constructive, which also provides an algorithm that computes an MMPE strategy profile in finite time (Readers who are interested in the proof can refer to \cite{eksin2017disease}). But unfortunately, even for such a simplified solution concept, no closed-form results in terms of expressing the MMPE action $a_i^*$ as a function of the current state $\bar{X}(t)$ is obtained. In the next subsection, we present several highlighted results obtained from simulations by the authors.

\subsubsection{Highlighted Results:}

\textit{A little empathy plays a huge role in bounding the basic reproduction number}: An important measure for an epidemic process is the basic reproduction number $R_0$, which measures the spread of an infectious disease from an initial sick individual in an otherwise susceptible host population. Whether $R_0>1$ or not is an indicator that the disease is likely to persist when there is a relatively low number of infected individuals. Hence, $R_0$ is an important measure relating the likelihood of disease persistence to network and utility weights. When individuals act according to an MMPE strategy profile, the following bound holds for $R_0$,
$$
R_0 \leq \frac{\beta}{\delta}\sum_{k=1}^{K} k P(k),
$$
where $P(k)$ is the proportion of individuals who have degree $k$ in the network, and $K \coloneqq \min\{ \lfloor c_1/c_2\rfloor, n\}$. Here, $n$ is the largest degree an individual has in the network. When there are no protective and preemptive measures taken, the bound for the network is $R_0 \leq \frac{\beta}{\delta}\sum_{k=1}^{n} k P(k)$.  If the empathy constant is close to zero such that $ c_2$ is smaller than $\frac{1}{n}c_0$, one recovers the bound for the contact network models with no protective or preemptive measures. When individuals weigh the risk of infecting others and the costs of taking measures equally, i.e., $c_0 =c_2$, the basic reproduction number $R_0$ is well bounded, i.e., $R_0\leq \frac{\beta}{\delta} P(1)$. In a scale-free network with degree distribution $P(k) \sim k^{-2}$, the bound becomes
\begin{equation}\label{eq:R0ScaleFreeBound}
R_0 \leq \frac{\beta}{\delta}  \frac{1}{2}(n+1).
\end{equation}
If the empathy weight $c_2$ is negligible ($c_2 < c_0/n$), meaning individuals do not care about infecting others, then the bound increases logarithmically with the size of the population, i.e., $R_0 \leq \frac{\beta}{\delta}  \frac{1}{2}(n+1)$. Indeed, the bound is the exact reproduction number for the contact network SIS model with no individual behavior response to disease prevalence \cite[Ch. 17]{newman2010networks}. The effect of individual response on $R_0$ appears as the empathy weight $c_2$ increases. To guarantee $R_0 <1$, the central authority needs to promote altruism among individuals to guarantee
\begin{equation}\label{eq:thresholdEmpatyWeight}
c_2 > \frac{c_0}{\exp(2\delta/\beta)-1}.
\end{equation}
To show the accuracy of the critical $c_2$ given by (\ref{eq:thresholdEmpatyWeight}), Eksin et al. compares the $R_0$ bound given by (\ref{eq:R0ScaleFreeBound}) to simulated $R_0$ value in \cite[Figure 4]{eksin2017disease}.

\textit{Risk aversion alone cannot eradicate the disease but it can affect the infection level}: While the risk aversion weight $c_1$ does not appear in any of the bounds for $R_0$,  $c_1$ cab affect the infection level of the population under the MMPE strategy even though it is not shown how $c_1$ affects the infection level. The authors show that when the empathy weight $c_2$ is zero, the outbreak threshold condition stays the same, i.e., $\beta \lambda_{max}(A)/\delta >1$ for any $c_1\in[0,1]$. This indicates that risk aversion alone cannot help eradicate the disease without the empathy of infectious individuals in this game.

\subsubsection{Discussions}

The game framework formulated here by Eksin et al. is similar to the fine-grained framework introduced in Section~\ref{subsec:FineFrainedFramwork}. Eksin et al. made two simplifications to ensure the analysis and the numerical computation are feasible. One is the use of MMPE as a solution concept to avoid resorting to dynamic programming techniques to solve the problem, which also facilitates the process of showing the existence of the equilibrium. Another one is the assumption that every individual knows the complete state information $\bar{X}(t)$ at every stage $t$.

Compared with \cite{reluga2010game} by T. Reluga, there are two differences in terms of the epidemic models. First, Eksin et al. utilize a networked model while T. Reluga chooses a shapeless epidemic model. Second, Eksin et al. select a stochastic model while T. Regula leverages a deterministic model. The networked epidemic model describes the health status at an individual level. This framework allows modeling the local interactions: protecting oneself and one's neighbors. However, the networked epidemic models limit the number of individuals in the population $N$ to a finite number. As $N$ increases, even the computation of a simplified equilibrium concept such as the MMPE becomes intractable. 

Eksin et al. have focused on investigating the effects of individual measures on disease eradication through the lens of the basic reproduction number $R_0$. Hence, the results here shed very little light on the effects of empathy and risk aversion on longer-term epidemic processes when the disease becomes endemic. Investing such individual measures in long-term spreading processes can be a future research direction. The authors do analyze the price of anarchy (PoA) under the MMPE and it is shown that the price of anarchy is bounded as
$$
1\geq \textrm{PoA} \geq 1- \frac{\max_{i\in\mathcal{N}}\vert\mathcal{N}_i\vert\max\{c_1,c_2\}}{n c_0}.
$$
This upper bound can be arbitrarily bad, i.e., in the order of $1/n$. That means without enforced measures from the central authority, an individual's response can be inefficient in terms of maximizing the social benefit. Hence, it is essential to couple the decentralized individual responses to disease prevalence with centralized policies such as public announcements, mass organized isolation, or vaccination campaigns arranged by the central authorities.

\subsection{Individual Weight Adaptation: An $N$-Player Differential Game \cite{huang2020differential}}\label{subsec:huang2020differential}

In \cite{huang2020differential}, Huang et al. proposes an $N$-player differential game framework based on the $N$-intertwined deterministic epidemic model (\ref{HeteroMieghemSIS}). The author mainly focuses on mitigating virus spreading on computer networks. But the modeling, the analysis, and the results of \cite{huang2020differential} can shed light on the epidemic spreading in the human population. In the differential game, each individual aims to strike a balance between being infected and loss of benefit (economic or social) by reducing his/her connection to neighbors. With a deterministic epidemic model, using Pontryagin's maximum principle, the authors are able to obtain structural results that help understand human behavior under an epidemic. The $N$-intertwined epidemic model is networked, allowing the authors to capture the local interactions among individuals and their neighbors. The authors also propose a penalty scheme, on behalf of the central authority, to achieve social welfare for the population. 

\subsubsection{Modelling}

\textit{Modelling interventions through a weighted network with time-varying weights}: A directed weighted network is a network where the directed connections among individuals have weights assigned to them. The weights between two individuals capture the intensity of the connectivity and the directions indicate the direction in which the disease or virus can spread. The weight between individual $i$ and individual $j$ at time $t$ is denoted by $w_{ij}(t)\in[0,1]$. In a directed network, $w_{ij}(t)$ is not necessarily equal to $w_{ji}(t)$ for any $t$. Individual $i$ can reduce his/her risk of contracting the virus by cutting down $w_{ij},j\in\mathcal{N}_j$, i.e., its connectivity with his/her neighbors. In a computer network, this can be done by reducing file transmission from one server to another or downloading fewer packages from some servers.

\textit{The epidemic model}: Huang et al. adopts the $N$-intertwined SIS epidemic model, in which the probability of individual $i$ being infected at time $t$ is denoted by $p_i(t)$. The evolution of $p_i$ is coupled with the states of other individuals in the network:
\begin{equation}\label{eq:epidemicmodelHuang}
\dot{p}_i = [1-p_i(t)]\beta \sum_{j\in \mathcal{N}_i^{out} }  w_{ij}(t) p_{j}(t) - \delta p_i(t),\ \ \ i=1,2,\cdots,N,
\end{equation}
where $\mathcal{N}_i^{out}$ denotes the set of neighbors individual $i$ reaches out for. If $p_j(t)$ is high, individual $i$ can reach out less to individual $j$ to avoid being infected, i.e., reducing $w_{ij}(t)$, and restoring its connection to individual $j$ after $p_j(t)$ decreases. The epidemic model (\ref{eq:epidemicmodelHuang}) is a variant of the $N$-intertwined deterministic model (\ref{MieghemSIS}) we have introduced in Section~\ref{subsubsec:DeterministicModels} with two differences. One is that here, huang et al. uses weights that can be adaptive to describe the network instead of fixed adjacency matrix elements. Another one is that the transmission is now directional. Individual $i$ being able to infect individual $j$ does not mean individual $j$ can be infected by individual $j$. Such cases appear more in computer networks. In human population, the transmission is often times mutual. Hence, it is more reasonable to use an undirected  network to model connections in human population.

\textit{The cost function:} The authors consider completely selfish individuals associating with two costs. One cost arises from infection captured by $f_i:[0,1]\rightarrow \mathbb{R}^+$. Another cost is from inefficiency or performance degradation induced by deviating from the original connections $a_{ij}$ for $i,j\in\mathcal{N}$, which is captured by a convex function $g_{ij}:\mathbb{R}\rightarrow \mathbb{R}^+$. Individual $i$ optimizes over a finite horizon $[0,T]$ with respect to its cost $J_i$:
\begin{equation}\label{eq:individualcostHuang}
J_i(p_i, p_{\mathcal{N}_i^{out}}, \{w_{ij}\}_{j\in\mathcal{N}_i^{out}}) = \int_0^T f_i(p_i(t)) + \sum_{j\in\mathcal{N}_i^{out}}g_{ij}(w_{ij}(t) - a_{ij})dt,
\end{equation}
where $a_{ij}$ is the original elements in the adjacency matrix. The original weights $a_{ij},i,j\in\mathcal{N}$ describe the connectivity before the epidemic hits. The original weights $a_{ij},i,j\in\mathcal{N}$ are assumed to be optimal for each individual meaning everyone keeps the best way of connecting to others for themselves. Any deviation from the original weights at time $t$ will induce inefficiency or performance degradation quantified by $\sum_{j\in\mathcal{N}_{i}^{out}} g_{ij}(w_{ij}(t) - a_{ij})$. The first term inside the integral of (\ref{eq:individualcostHuang}) represents individual $i$'s cost of infection. Different from the work by Eksin et al. \cite{eksin2017disease}, the individuals here are completely selfish who only care about their own infection and inefficiency or performance degradation. However, the goal of the central authority is to minimize the aggregated social cost
\begin{equation}
J_c = \int_0^T \sum_{i\in\mathcal{N}} f_i(p_i(t)) +\sum_{i\in\mathcal{N}} \sum_{j\in\mathcal{N}_i^{out}}g_{ij}(w_{ij}(t) - a_{ij})dt.
\end{equation}
The terminal time $T$ is the time when the mass installation of an anti-virus or anti-malware patch is implemented.

\subsubsection{Analysis}

The authors apply Pontryagin's minimum principle to obtain a Nash equilibrium for the $N$-person differential game defined by (\ref{eq:epidemicmodelHuang}) and (\ref{eq:individualcostHuang}). Under the Nash equilibrium strategy, individuals adapt their weights according to the following rule.
\begin{theorem}
The Nash equilibrium strategy profile $\{w_{ij}^*\}_{i\in\mathcal{N},j\in\mathcal{N}_i}$ for the $N$-person differential game defined by (\ref{eq:epidemicmodelHuang}) and (\ref{eq:individualcostHuang}) are adapted according to
\begin{equation}\label{eq:NEAdaptationHuang}
w_{ij}^*(t) = \begin{cases}
0,\ \ \ & -\phi_{ij} \leq g'_{ij}(-a_{ij}),\\
(g_{ij}')(-\phi_{ij}(t)),\ \ \ &g_{ij}'(-a_{ij}) < -\phi_{ij}(t) < g_{ij}'(0),\\
a_{ij},\ \ \ &-\phi_{ij}(t)\geq g_{ij}'(0),
\end{cases}
\end{equation}
for $i\in\mathcal{N},j\in \mathcal{N}_i^{out}$, where $\phi_{ij}(t)\coloneqq x_{ii}(t) (1- p_i^*(t))\beta_j p_j^*(t)$. Here, $x_{ii}(\cdot)$ is the $i$-th component of the costate function $\mathbf{x}_i (\cdot)$ whose dynamics are governed by 
$$
\mathbf{x}_i(t) = \Lambda_i (t,\{p_i(t)\}_{i\in\mathcal{N}}, \{w^*_{ij}(t)\}_{i\in\mathcal{N},j\in\mathcal{N}_i}) \mathbf{x}_i(t) + \lambda_i,\ \ \  i\in\mathcal{N},
$$
where $\lambda_i= - df_i/d p_i$ and $\Lambda$ is a matrix defined as
$$
\Lambda_{i,mn} = \begin{cases}
\sum_{j\in\mathcal{N}_m^{out}} w^*_{mj}\beta_j p_j^* + \delta_m,\ \ \ &\textrm{if }n=m,\\
-(1-p^*_n(t))w_{nm}^*(t)\beta_m,\ \ \ &\textrm{if }n\in\mathcal{N}_m^{in},\\
0,\ \ \ &\textrm{otherwise}.
\end{cases}
$$
\end{theorem}

One can observe from (\ref{eq:NEAdaptationHuang}) that the weight adaptation $w_{ij}^*$ under Nash equilibrium depends on $\phi_{ij}(t)\coloneqq x_{ii}(t) (1- p_i^*(t))\beta_j p_j^*(t)$. That means the weight adaptation $w_{ij}^*$ is decided by the costate component $x_{ii}$, which carries the information about the whole population, individual $i$'s health status $p_i$, and his/her neighbor's health status. To see the dependence more clearly, one can assume that $g_{ij}$ is quadratic, i.e., $g_{ij}(x) = \frac{1}{2} x^2$. Then, 
\begin{equation}\label{eq:NEAdaptationQuadraticHuang}
w_{ij}^*(t) = \begin{cases}
0,\ \ \ & \phi_{ij} \geq a_{ij},\\
-\phi_{ij}(t),\ \ \ &a_{ij} > \phi_{ij}(t) > 0,\\
a_{ij},\ \ \ &\phi_{ij}(t) \leq 0.
\end{cases}
\end{equation}
One can draw several intuitive conclusions immediately from (\ref{eq:NEAdaptationHuang}) and (\ref{eq:NEAdaptationQuadraticHuang}). The first is that individual $i$ reaches out less to his/her neighbor $j$ if the neighbor is infected with high probability, i.e., $w_{ij}^*$ decreases when $p_j^*(t)$ increases. The second is that individuals tend to reach out less to neighbors who can get infected easily, i.e., a high $\beta_j$ leads to a lower $w_{ij}^*$.

The weight adaptation $w_{ij}^*(t)$ is continuous over time if $g_{ij}$ is convex. If $g_{ij}$ is concave, individuals would implement a bang-bang type of control:
$$
w_{ij}^*(t) = \begin{cases}
0,\ \ \ &\phi_{ij}(t) \geq \frac{g_{ij}(-a_{ij})}{a_{ij}},\\
a_{ij},\ \ \ &\phi_{ij}(t) > \frac{g_{ij}(-a_{ij})}{a_{ij}}.
\end{cases}
$$
Different from the goals of individuals, the central authority aims to minimize social costs $J_c = \sum_{i\in\mathcal{N}} J_i$. Huang et al. show that if the central authority applies a penalty $c_i(t) = \sum_{j\in \mathcal{R}_i}f_j(x_j)$ to individual $i$ at each time $t$, the central authority can achieve social social optimum without enforcing weight adaptation rules for every individual. Here, $\mathcal{R}_i$ is the set of individuals that can be reached by individual $i$.

\subsubsection{Results}

\textit{Not just current health status matters, inherent vulnerability also matters}: From 
(\ref{eq:NEAdaptationHuang}) and (\ref{eq:NEAdaptationQuadraticHuang}), we know that individuals cut down their connections with individuals who are currently infected. Beyond that, individuals also lower more weight of connections with individuals who are inherently more vulnerable to the virus even though they may not be infected currently. For example, in computer network,  one may reduce data transfer from a server whose security level is very low.

\textit{The `what-the-heck' mentality}: The `what-the-heck' mentality is one that when something undesired happens, one gives up making effort to fix it. Similar mentality also appears in results of (\ref{eq:NEAdaptationHuang}) and (\ref{eq:NEAdaptationQuadraticHuang}). The term $(1-p_i(t))$ in $\phi_{ij}$ suggests that if individual $i$ is infected, he/she just restores all his/her connections with his/her neighbors. Since the network studied here by Huang et al. is directed, infected individuals restoring their connections will not cause more infection. It is worth investigating whether this result still holds or not in undirected networks. If it holds, this uncaring behavior may lead to a higher infection level in the population.

\textit{The weight reduction lasts till the last minute}: The authors show that for all $i\in\mathcal{N}$, $x_{ii}(t)\geq 0$ for all $t\in[0,T]$ with $x_{ii}(t)=0$ if and only $t=T$. It is also shown that for every $i\in\mathcal{N}$, $p_{i}>0$ for all $t\in[0,T]$. That means every individual would more or less cut down his/her connections with neighbors during the spreading process and only restore his/her connections when mass installation of security patches or mass vaccination is available, i.e., $w^*_{ij}(t)<a_{ij}$ for $t\in[0,T)$ and $w_{ij}^*(T)=a_{ij}$. 

\textit{Equilibrium weight adaption outperforms social optimal weight adaptation in terms of lowering infection level of the whole network}: Using numerical examples, Huang et al. show in Figure~\ref{fig:InfectionLevelHuang} that even though the game-based scheme is inefficient in terms of minimizing the total cost $J_c$, it outperforms the optimal control-based scheme as we can observe that the infection level under the game-based scheme is always lower than the infection level under the optimal control based scheme. No matter what value $\alpha$ is, the equilibrium weight adaption scheme tends to mitigate the spreading of the virus better than the scheme without adaptation.

\begin{figure}
  \includegraphics[width=0.8\textwidth]{infectionlevelHuang.jpg}
\caption{\cite{huang2020differential} The dynamics of the whole network’s infection under the differential game-based weight adaptation scheme, the optimal control-based weight adaption scheme, and the scheme without weight adaptation when $f_i(p_i(t)) = \alpha p_i(t)$.} 
\label{fig:InfectionLevelHuang}     
\end{figure}

\subsubsection{Discussions}

The game framework formulated here by Huang et al. is, similar to the framework proposed by Eksin et al. in \cite{eksin2017disease}, built on a networked epidemic model. This framework allows the authors to capture the local interaction and the individual-level virus spreading. The authors here choose the deterministic epidemic models allowing them to obtain analytical results in terms of the Nash equilibrium of a finite horizon problem, while in \cite{eksin2017disease}, only the solution for the myopic Markovian Nash equilibrium can be computed numerically. However, since the $N$-intertwined deterministic epidemic model is an approximation of the stochastic model and the accuracy of the approximation is only guaranteed under certain conditions \cite{van2008virus}, it is not clear whether the game-theoretic model built upon the approximated spreading process provides an accurate approximation of human behavior. In general, the analysis of whether the equilibrium obtained under the approximated spreading process stays close to the equilibrium obtained under the stochastic process requires significant technical leaps, and it is a topic of active research. 

The Nash equilibrium obtained by the authors is an open-loop solution.  It is pointed out by \cite{khouzani2011saddle} that this solution is appropriate in the context of the security in networks, as the instantaneous state of each individual is impossible (or very costly) to follow. But it is still worth studying whether the closed-loop Nash equilibrium and the open-loop one produce the same value of the game and whether the open-loop strategy can be implemented in a feedback way in future endeavors. Another direction of research for this type of framework is to study the long-term behavior of individuals, i.e., the case when $T$ is large or infinite.

\section{{Game Theory on the Move and Potential Directions}}\label{sec:FutureEnvision}

In the previous section, we presented an overview of the current state of the art involving the game-theoretic models to study human behavior amid epidemics. {Game theory has catched many eyes of people who experiences the covid-19 pandemic and influenced policies both directly and indirectly. Since the prevalence of the COVID-19 pandemic, game theory has been on the move from academia to media and the government. The New York Times, a newspaper that can reach tens of millions readers worldwide, has published an article about how game theory can explain the waves we observed from the COVID-19 infection data, how game theory can help prioritize vaccines, and how game theory incorporated with COVID data can help make predictions \cite{Roberts2020}. The Forbes Magazine has featured an article about how a game theory-based model can enhance stock management of PPE (Personal Protective Equipment) supply in the hospitals of the English National Health Service (NHS) \cite{Varanasi21}. The Fortune Magazine has featured a commentary on how game theory can solve the vaccine rollout puzzle \cite{muggy2021} and help state authorities to minimize the cumulative distance traveled and congestion at facilities, or maximize efficient and equitable access to vaccines. The game theory-based approach for vaccine rollout planning has been proven valuable in after-the-fact analyses of the H1N1 vaccination campaign in 2009 \cite{stamm2017quantifying} and the response to Haiti’s cholera epidemic in 2010 \cite{muggy2020decentralized}.}

{Apart from media coverage, stakeholders such as the government and the US army are actively looking for expert advice from game-theoretic perspectives. The Ontario government in Canada has funded Chris Bauch's research project that studies how to re-open Ontario's economy without causing a resurgence by leveraging game theory-based models \cite{Ontario2020}. The Ontario government is leveraging the research results to prevent, detect and treat COVID-19. Amid the COVID-19 pandemic, the US Department of Defense (DoD) has awarded the University of Michigan in Ann Arbor \$6.5 million to study how officials at different levels of government can work together to maximize COVID-19 safe behavior \cite{DoD2020}. Beyond that, research results regarding the blood shortages during the COVID-19 pandemic has reached to the Admiral Brett Giroir of the US Department of Healthy \& Human Services in a letter sent from California Attorney General Xavier Becerra and signed by Attorney Generals of 21 other states \cite{Johnson2020}. The letter cited and demonstrated the research results in which Dr. Anna Nagurney has investigated the supply chain network competition among blood service organizations using a generalized Nash equilibrium framework and how it causes the blood shortages during the COVID-19 pandemic \cite{nagurney2019supply}. The idea of game theory also influences contact tracing application developers. Inspired by game theory and behavioral science, app developers develop an app called NOVID which uses an approach called ``inverts the incentives''. The app warns people ahead of time that danger was near instead of telling people they’d been exposed and, for other people’s good, quarantining themselves to stop further spread \cite{Samorodnitsky2021}.
}

Despite of the current impact that the game theory-based model made, there are still many hollows and research gaps in the  results presented that need to be well taken care of. This section spotlights several of the main research challenges, to what extent they are currently being tackled, and what needs to be done.

\subsection{The Justification of Game-Theoretic Methods for Deterministic Models}

Many game-theoretic frameworks have adopted continuous deterministic epidemic models since these models are easy to analyze and more likely to produce meaningful analytical results. Continuous-time deterministic models are  approximations of the epidemic spreading processes in the form of ODEs. The justification of passing from the original stochastic models to a continuous deterministic model in the presence of strategic interventions is barely mentioned in most papers. For example, it is an open problem of how the weight adaptation policies found for the $N$-intertwined deterministic model (\ref{eq:epidemicmodelHuang}) in Section~\ref{subsec:huang2020differential} relate to the policies found for its stochastic counterpart (\ref{2NExactSISModel}), and how the social distancing policies found in Section~\ref{subsec:reluga2010game} relate to the policies we would find for the stochastic model (\ref{StochasticPopModel}). 

Gast et al. has proposed a mean-field Markov decision process framework, which provides a rigorous justification of the use of a continuous optimal control approach for the virus spreading problem, and shows that the continuous limits provide insights on the structure of the optimal behavior for the discrete stochastic epidemic model \cite{gast2012mean}. The work by Gast et al. has focused on centralized problems  (i.e., optimal control problems). Tembine et al. have investigated the asymptotic behavior of a Markov decision evolutionary games as the size of the population grows to infinity \cite{tembine2009mean}. It is an open research question whether these mean-field results would hold for popular continuous epidemic models when game-theoretic decision-making is involved.

\subsection{The Role of the Central Authority: Mechanism Design and Information Design}

Recognizing the necessity of decentralized decision-making for controlling epidemics, a growing number of researchers have started to pay attention to game-theoretic decentralized solutions. However, most studies of this kind focused on the interactions among individuals \cite{bauch2004vaccination,bhattacharyya2010game,breban2007mean,chang2020game,dashtbali2020optimal,hayel2017epidemic,hota2020impacts,hota2016interdependent,hota2019game,ibuka2014free,lagos2020games,liu2012impact,reluga2010game,reluga2006evolving,zhang2013braess,eksin2017disease,eksin2019control,chapman2012using,adiga2016delay,huang2020differential,li2017minimizing,saha2014equilibria,trajanovski2015decentralized,trajanovski2017designing}. Some studies investigated the inefficiency of selfish decision-making demonstrated by the price of anarchy. Among these studies, some has not considered the role of the central authority \cite{reluga2006evolving,trajanovski2015decentralized}. Indeed, the central authority plays a significant role in controlling the epidemic spreading even when the interventions are not enforceable among individuals. When individuals make their own decisions to maximize their payoffs, it is important to consider mechanism design \cite{fudenberg1991game} and/or information design \cite{zhang2021informational} to create incentives for individuals and to achieve a certain degree of social optimum.

A handful of papers have studied the role of the central authority in an epidemic when individuals care for their payoffs \cite{aurell2020optimal,pejo2020corona}.  \cite{aurell2020optimal} proposed a Stackelberg game framework in which individuals play a non-cooperative game, and the central authority influences the nature of the resulting Nash equilibrium through incentives so as to optimize its own objective. In \cite{pejo2020corona}, the authors promote a mechanism design approach, on behalf of the central authority, weighing response costs vs. the social good. Some researchers have studied how information can affect human behavior amid an epidemic \cite{wang2020epidemic,lagos2020games}. So far, how should central authority disseminate information to help individuals fight against the virus remains an open problem.

\subsection{Information Matters}

Most papers assume complete information in their game-theoretic frameworks \cite{reluga2006evolving,reluga2010game,huang2019achieving,huang2020differential,eksin2017disease,eksin2019control}. The assumption of complete information in the context of an epidemic study is that every individual knows, when they make decisions, the health status of every other individual in the network, which is unrealistic and impractical. Hence, dynamic games with partial information or incomplete information need to be applied to studying human behavior amid an epidemic. However, {papers} following this direction of research are rare.

Obtaining information can be costly. For example, if an individual needs to find out whether he/she is infected or not, he/she has to pay for a testing kit or spend time commuting to a testing spot. If the central authority wants to find out the infection level of the whole population, it needs to deploy a large number of resources to conduct mass testing or mass survey. How decision-makers decide when to obtain information and how frequently information needs to be updated become interesting questions worth addressing. In \cite{theodorakopoulos2012selfish}, the authors revealed a counterintuitive fact that the equilibrium level of infection increases as the users’ learning rate increases. That means the more frequently the information exchanges, the higher the infection level at the equilibrium. In \cite{huang2019continuous}, the authors consider costly information in a continuous Markov decision process, and the authors show it is not necessary to obtain information very often to maintain satisfactory performance. Understanding the right timing to conduct mass testing to help all decision-makers resist the diseases is essential, especially for those long-lasting low-infection-level diseases. However, this problem has not been touched in the literature yet.

\subsection{Human are not Completely Rational}
As with all game-theoretic models, human behavior is unlikely to completely agree with the equilibria due to the lack of information or solid prior beliefs. Nevertheless, very few papers have studied the effect of bounded rationality of individuals. Recent studies have included irrationality in evolutionary games when individuals are not entirely  rational \cite{poletti2012risk}. Advanced solution concepts such as the `Quantal Response Equilibrium' have been studied to account for bounded rationality of individuals \cite{goeree2010quantal}. In \cite{hota2019game}, Hota et al. pointed out that humans often perceive probabilities differently from their true values, and they modeled the misperception of the infection risk among individuals. However, to understand the bounded rationality of human behavior, more studies need to be carried out in the future.

\section{Conclusion}

This article has provided a tutorial combined with a review about the integration of decision models into epidemic models. We focus on how game-theoretic models are coupled with epidemic models to understand the human-in-the-loop epidemic spreading process. Targeting at readers from the game theory community and the dynamic game community, this review focuses more on introducing the popular epidemic models, the pros and cons of different epidemic models, and how interventions can be modeled in the epidemic models rather than the basics of game theory. We provide a taxonomy of game-theoretic framework for human-in-the-loop epidemic modeling. From the taxonomy, game theorists from different branches can identify topics that are related to them the most. Although this review focuses on disease and epidemics in the human population, similar mathematical formulations, tools, and results can extend directly to many different spreading processes including malware spreading, information dissemination, etc.

In this review, we strive to provide  a comprehensive view of the literature. Game-theoretic modeling for epidemics is an emerging topic. There are many useful  and important game-theoretic frameworks in the literature that have not been included. We encourage interested readers to keep track of the current of the art. Despite the vast amount of papers studying the game-theoretic modeling for epidemics, there are still many open questions and gaps waiting to be addressed and filled, especially the ones we introduced in Section~\ref{sec:FutureEnvision}. We believe game theorists and dynamic game theorists can play a significant role, address these questions, and fill the  void in theory and  practice. Epidemics have taken and are taking a devastating toll on human society. The game-theoretic framework for understanding the human-in-the-loop epidemic spreading will make a real societal impact.




%
%

\bibliographystyle{spmpsci}      
\bibliography{refs.bib}   

%
%

\end{document}